\documentclass{aa}  

\usepackage{graphicx}
\usepackage{txfonts}
\usepackage{natbib,twoopt}
\usepackage{hyperref}
\bibpunct{(}{)}{;}{a}{}{,} 
\makeatletter
\newcommandtwoopt{\citeads}[3][][]{\href{http://adsabs.harvard.edu/abs/#3}%
{\def\hyper@linkstart##1##2{}%
\let\hyper@linkend\@empty\citealp[#1][#2]{#3}}}
\newcommandtwoopt{\citepads}[3][][]{\href{http://adsabs.harvard.edu/abs/#3}%
{\def\hyper@linkstart##1##2{}%
\let\hyper@linkend\@empty\citep[#1][#2]{#3}}}
\newcommandtwoopt{\citetads}[3][][]{\href{http://adsabs.harvard.edu/abs/#3}%
{\def\hyper@linkstart##1##2{}%
\let\hyper@linkend\@empty\citet[#1][#2]{#3}}}
\newcommandtwoopt{\citeyearads}[3][][]%
{\href{http://adsabs.harvard.edu/abs/#3}
{\def\hyper@linkstart##1##2{}%
\let\hyper@linkend\@empty\citeyear[#1][#2]{#3}}}\makeatother

\begin{document}

   \title{Activity cycles in members of young loose stellar associations}

   \author{E. Distefano
          \inst{1}
          \and
          A. C. Lanzafame
          \inst{2}
          \and
          A. F. Lanza\inst{1}
          \and S. Messina\inst{1}
          \and F. Spada\inst{3}
         }

   \institute{ INAF - Osservatorio Astrofisico di Catania\\
              Via S. Sofia, 78, 95123, Catania, Italy\\
              \email{elisa.distefano@oact.inaf.it}
            \and University of Catania, Astrophysics Section, Dept. of Physics and Astronomy\\
          Via S. Sofia, 78, 95123, Catania, Italy  
          \and Leibniz-Institut f\"ur Astrophysik Potsdam (AIP)\\
           An der Sternwarte 16, D-14482, Potsdam, Germany \\         
             }

   \date{Received ; accepted }

 
  \abstract
   {Magnetic cycles analogous to the solar one have been detected in tens of solar-like stars by analyzing long-term time-series of different magnetic activity indexes. The relationship between the cycle properties and global stellar parameters is not  fully understood yet. One reason for this is the sparseness of the data.}  
   {In the present paper we searched for activity cycles in a sample of 90 young solar-like stars with ages  between 4 and 95 Myr with the aim to investigate the properties of activity cycles in this age range.  }
   { We measured the  length $P_{\rm cyc}$ of a given cycle by analyzing the long-term time-series of three different activity indexes: the period of rotational modulation, the amplitude of the rotational modulation and the  median magnitude in the V band. For each star, we computed also the global magnetic activity index <IQR> that is proportional to the amplitude of the rotational modulation and can be regarded as a proxy of the mean level of the surface magnetic activity.
 }
   {We  detected activity cycles in 67 stars. Secondary cycles were also detected in 32 stars of the sample. The lack of correlation between $P_{\rm cyc}$ and $P_{\rm rot}$ and the position of our targets in the $P_{\rm cyc}/P_{\rm rot} -Ro^{-1}$ diagram suggest that these stars belong to the so-called Transitional Branch and that the dynamo acting in these stars is  different from the solar one and from that acting in the older Mt. Wilson stars. This statement is also supported by the analysis of the butterfly diagrams whose patterns are very different from those seen in the solar case.  
   We computed the Spearman correlation coefficient $r_{\rm S}$ between $P_{\rm cyc}$, <IQR> and different stellar parameters. We found that $P_{\rm cyc}$ in our sample is uncorrelated with  all the investigated parameters. The <IQR> index is positively correlated with the convective turn-over time-scale, the magnetic diffusivity time-scale $\tau_{\rm diff}$, and the dynamo number $D_{\rm N}$, whereas it is anti-correlated with the effective temperature $T_{\rm eff}$, the photometric shear $\Delta\Omega_{\rm phot}$ and the radius $R_{\rm C}$ at which the convective zone is located. 
   We investigated how $P_{\rm cyc}$ and <IQR> evolve with the stellar age.
   We found that $P_{\rm cyc}$ is about constant and that <IQR> decreases with the stellare age in the range 4-95 Myr.
Finally we investigated the magnetic activity of the star AB Dor A by merging  ASAS time-series with previous long-term photometric data. We estimated the length of the AB Dor A primary cycle as $P_{\rm cyc} = 16.78 \pm 2 \rm yr$  and we found also shorter secondary cycles with lengths of 400 d, 190 d  and 90 d respectively. } 
{}


   \maketitle
%
\section{Introduction}

The present paper addresses the occurrence and properties of magnetic activity cycles in young solar-like stars with ages spanning the range $4-95~\rm Myr$.
The study of activity cycles requires the availability of long-term time-series of one or more magnetic activity indexes.

The Solar magnetic activity has  been well monitored in time by recording the
 measurements of different activity indexes such as the Total Solar Irradiance (TSI), the sunspots number, the flare occurrence rate, and the intensity of specific spectral lines.
The analysis of long-term time-series of these activity proxies revealed that the Sun exhibits periodic or quasi-periodic variability phenomena occurring over different time-scales ranging from 30~d to 80~yr.
The 30-d periodicity is induced by the solar rotation that modulates the visibility of spots and faculae  over the solar disk. 
A 11-yr period, the so-called Schwabe cycle, is related to the evolution of the solar magnetic field and is associated with a cyclic variation of the sunspots number and of the average latitude at which spots and faculae occur.

The level of the solar magnetic activity is a complex function of time and other signals are superimposed over the 11-yr cycle.

A 154-d periodicity, for instance, was detected in $\gamma$-ray activity \citep{1984Natur.312..623R}, in the  Mt. Wilson sunspot index \citep{2002ApJ...566..505B}, and in the sunspot area \citep{1990ApJ...363..718L,1998Natur.394..552O}.
These 154-d cycles are usually named "Rieger cycles" because they were detected, for the first time, by \cite{1984Natur.312..623R}.

The expression "quasi-biennial variations" is instead used to refer to periodic or quasi-periodic phenomena that occur with time-scales ranging from 1 yr to 3 yr.
\cite{1967SoPh....1..107G,1977SoPh...51..175G}, for instance, analyzed  variations in sunspots numbers, large flares occurrence rate and coronal green-line emission and found that each solar cycle exhibits two maxima separated by 2-3 yr  intervals.
Variability phenomena occurring in a 2-yr time-scale have also been  detected in the 35-years TSI  time-series analyzed by  \cite{2015A&A...583A.134F}. 
A review of the quasi-biennial variations detected in solar activity and of the physical processes invoked to explain such variations can be found in \cite{2015LRSP...12....4H}.
Finally, on longer time-scales, the Sun activity is characterized by the 22-yr Hale cycle \citep{1919ApJ....49..153H} associated with a reverse in the sunspots polarity and the 80-yr Gleisseberg cycle \citep{Gleissberg} that is a a cyclic modulation of the amplitude of the 11-yr cycles.

Stars with a spectral type later than F5 have a magnetic activity similar to that observed in the Sun and are therefore characterized by activity cycles similar to those occurring in our star.
The analysis of long-term photometric time-series allowed the detection of activity cycles in a wide sample of late-type stars. The first studies on stellar activity cycles were based on the results of the Mount Wilson Ca II H\&K survey (Wilson 1968) and were conducted by \cite{1978ApJ...226..379W}, \cite{1981PASP...93..537D}, \cite{1985ARA&A..23..379B} and \cite{1995ApJ...438..269B}. These authors measured the rotation period $P_{\rm rot}$ and the cycle length $P_{\rm cyc}$ for tens of stars by analyzing the temporal variations of the photometric fluxes  in two 0.1 nm passbands centered on the core of the Ca II H and K emission lines. 
Recently, \cite{2016A&A...590A.133O} analyzed the full  Ca II H\&K time-series collected at Mount Wilson for a sample of 29 stars. Their work exploits 36-year time-series that are ten-years longer than those analyzed by  \cite{1995ApJ...438..269B} and allow a better frequency resolution. 

\cite{2002A&A...393..225M} exploited long-term photometry in the Johnson V band to measure the rotation period $P_{\rm rot}$ and the length of activity cycle $P_{\rm cyc}$ in a sample of six young solar analogues. \cite{2009A&A...501..703O} analyzed long-term spectroscopic and photometric observations to study the activity cycles in a sample of 20 late type-stars.

\cite{2016A&A...588A..38L} studied activity trends in a sample of 21 young solar-type stars by analyzing long-term photometry collected in Johnson B and V bands.
\cite{2016A&A...595A..12S} detected long-term activity cycles in a sample of 49 nearby main-sequence stars by analyzing the V-band time-series collected by the ASAS survey \citepads[All Sky Automatic Survey,][]{1997AcA....47..467P}.
Recently, high precision photometric time-series  acquired by space-borne telescopes were used to study stellar cycles. \cite{2014MNRAS.441.2744V}, for instance, exploited Kepler data to study activity cycles in 39 fast-rotating late-type stars and   \cite{2015A&A...583A.134F} studied activity cycles in 16 late-type stars by analyzing CoRoT time-series.

Since stellar activity cycles have been detected, a link has been searched  between the  cycle length $P_{\rm cyc}$, the magnetic activity indexes and global stellar parameters like the rotation period $P_{\rm rot}$, the Rossby Number $Ro$ and the age \citep[see e.g.][and references therein]{2016MNRAS.462.4442S,2016A&A...590A.133O,2014A&A...572A..34G}. These relationships are important to probe the theoretical models developed to  simulate the magnetic dynamos. 
The data currently available are still incomplete especially for young solar-like stars and the picture of the relationships between magnetic activity and global stellar parameters is far from being clear.

The relationship between the parameters $P_{\rm cyc}$ and $P_{\rm rot}$ is, for instance, not fully understood. Several authors found that  $P_{\rm cyc}$ and $P_{\rm rot}$  are correlated with each other \citep[see e.g.][]{1996ApJ...460..848B,2007ApJ...657..486B,2009A&A...501..703O,2016A&A...590A.133O}.
Other authors found no correlation between the two parameters \citep{2016A&A...588A..38L}.
Other works found a  correlation in main-sequence F, G and K stars and no correlation in M type stars \citep{2012ARep...56..716S,2014MNRAS.441.2744V,2016A&A...595A..12S}. This last feature has been interpreted as a difference in the dynamo mechanisms acting in FGK stars and M stars.

The study   of the relationship between the $\frac{P_{\rm cyc}}{P_{\rm rot}}$ ratio and the Rossby number showed that  stars seem to lie in different activity branches \citep[see e.g.][]{1998ApJ...498L..51B,2016A&A...588A..38L} possibly connected to different kinds of dynamos.
\cite{2007ApJ...657..486B}  remarked that the Sun has an anomalous position with respect to these branches. It is unclear if the Sun is a special case or if other stars exhibit a similar behavior.

The evolution of the $\frac{P_{\rm cyc}}{P_{\rm rot}}$ ratio with the stellar age has been studied by 
\cite{1993ApJ...414L..33S} and, more recently, by \cite{2016A&A...590A.133O} in the age range 200-6200 Myr.

In the present work we searched for activity cycles in a sample of 90 late-type stars belonging to young loose stellar associations. 
 In \cite{2016A&A...591A..43D} (hereafter Paper I) we investigated the rotational properties of these stars and we determined a lower limit for their SDR (Surface Differential Rotation). In the present paper we focused in their magnetic activity. As, remarked in Paper I,  the ages and the spectral types of these stars span the ranges 4-95 Myrs and G5-M4, respectively.
This means that our sample comprises both fully convective stars as well as stars with a radiative core surrounded by a convective envelope.
Hence the data used here allow the characterization of activity cycles in stars with different structures and ages and to extend the age range investigated by  \cite{1993ApJ...414L..33S} and   \cite{2016A&A...590A.133O}.

The data used to perform our analysis and the procedure followed to measure the length of activity cycles are described in Sec. 2 and in Sec. 3, respectively.  The results of our analysis are reported in Sec. 4 and discussed  in Sec. 5.  In Sec. 6, the conclusions are drawn.

\section{Data}
\subsection{Targets}
 In this paper we exploited long-term photometric time-series collected by the ASAS  survey \citepads[All Sky Automatic Survey,][]{1997AcA....47..467P} and we searched for activity cycles in a sample of 90 solar-like stars belonging to young loose stellar associations. These targets  are listed in Table \ref{parameters} together with their main astrophysical parameters.
In Table \ref{association} we reported, for each association, the number of stars investigated  and the number of stars in which we were able to detect one or more activity cycles. 
The same ASAS time-series have been exploited in Paper I to investigate the rotational properties of our target stars.
 
Note that in Paper I we also investigated 22 stars for which the Super-WASP \citep{2010A2006PASP..118.1407ParchiveA...520L..10B} time-series were available. These data have a photometric accuracy better than ASAS data. However, a typical Super-WASP datasets spans only a 4-year interval and is characterized by  observation gaps of several months. The ASAS time-series, despite their lower photometric precision, are best suited to study long-term variability. Indeed these time-series span an interval of about 9 years and are characterized by shorter observation gaps due to the use of two different observing stations located at Las Campanas Observatory, Chile and at Haleakala, Maui, Hawaii, respectively. For this reason we decided to analyze  the ASAS data only.

\subsection{Magnetic activity indexes}
All works cited in the introduction measured the lengths of activity cycles by searching for a periodicity in long-term time-series of different activity indexes.

In the present paper, we exploited the long-term time-series of three   activity indexes that have been obtained from the analysis of the photometric data collected by the ASAS  survey. 
In  Paper I, we processed the ASAS data with a sliding window algorithm that divided each photometric time-series in segments of a given length T. For each segment, we computed three activity indexes: the  rotation period $P_{\rm rot}$, inferred by  the rotational modulation of the light curve by magnetically Active Regions (ARs), the median magnitude ($V_{\rm med}$), and the Inter-Quartile Range (IQR)  defined as the difference between the 75-th  and the 25-th percentiles of V magnitude values (different examples of these activity indexes time-series can be seen in Figs. 2-5 of Paper I and in Figs. 1-4, 12-13 of the present work.)

In Paper I we processed the ASAS photometric time-series by using three different values of the sliding-window length, i.e. T=50 d, 100 d and 150 d, producing three different time-series for each activity index. The results given in Paper I have been inferred by the time-series obtained with T=50 d. As discussed there, the use of  sliding-window lengths of T =100 d and T = 150 d tends to filter out the variability phenomena occurring over a time-scale shorter than 100 d, flattening the amplitude variations of the different activity indexes (see. Figs. 2-5 of Paper I).
 However, the activity indexes time-series obtained with T=100 and 150 d are more regular and smoother than those obtained with T=50 d and are more suitable to detect long-term variations due to activity cycles.
Therefore, while T=50 d is an optimal choice for analyzing SDR as in Paper I, longer segments, e.g. T=100 d as used here, are better suited to search for activity cycles. This difference in T obviously does not introduce any inconsistencies in, e.g., correlation analysis between activity cycles and SDR.

In Paper I, we showed that the rotation period detected in different segments slightly changes in time. This trend could be induced by the combined effect of  stellar Surface Differential Rotation (SDR) and of  ARs  migration in latitude.
Indeed  ARs placed at different latitudes rotate with different frequencies in case of SDR. The ARs migration, due to a Schwabe-like cycle, induces a temporal variation in the detected photometric period $P_{\rm rot}$.  
Such a parameter can be therefore regarded as a tracer of the spot latitude migration.

Note that this activity index has to be treated with caution. \cite{2016A&A...588A..38L} and \cite{2016A&A...592A.140L} noticed that the different rotation periods detected in different  segments could be also a consequence of the ARs growth and decay occurring simultaneously at various latitudes and longitudes on the star. However, as remarked by \cite{2016A&A...588A..38L},  the use of segments with a  short duration should limit this effect (see the discussion on the sliding-window length in Sec. 3.1 of Paper I and in Sec. 4.1 of \cite{2016A&A...588A..38L} for more details).
This activity index has been recently used by  \cite{2014MNRAS.441.2744V} to investigate stellar cycles on Kepler data.

The $V_{med}$ index is related to the axisymmetric part of the spot distribution on the stellar surface. In fact, the higher  the fractional area covered by spots evenly distributed in longitude, the lower  the average photometric flux in the V band,. 
This is one of the most used activity index to investigate activity cycles \citep[see e.g.][]{2000A&A...358..624R,2002A&A...393..225M,2016A&A...588A..38L}

The IQR index is proportional to the stellar variability amplitude. The variability amplitude is also an index widely used to study activity cycles \citep[see e.g.][]{2000A&A...358..624R,2015A&A...583A.134F,2016A&A...588A..38L}. Note that the variability of solar-like stars is due to different phenomena acting in different time-scales like the intrinsic evolution of spots and faculae, the intrinsic evolution of ARs complexes and the flux rotational modulation induced by ARs. In this paper, the IQR index was computed in 100-d long segments. In these intervals the main source of variability is the rotational modulation \citep[see e.g.][]{2002A&A...393..225M,2003A&A...403.1135L}. Hence the IQR index is proportional to the amplitude of the rotational modulation signal and is related to the non-axisymmetric part of the spot distribution. 

Finally, for each star we computed also a global magnetic index that is given by the average value <IQR> of the IQR time-series.  The <IQR> values are reported in Table \ref{parameters} for all the stars investigated here.

\begin{table*}
\caption{Stellar associations investigated in the present work. }
\label{association}
\centering
\begin{tabular}{llll}
\hline
Association & age    & $N_{stars}$     & $N_{cyc}$   \\
                   &   Myr      &   &           \\
\hline
$\epsilon$ Chamaleontis ($\epsilon$ Cha) & 3-5  & 9 & 6 \\
$\eta$ Chamaleontis ($\eta$ Cha) &  6-10  & 4 & 4 \\
TW Hydrae (TWA) &  8-12 &  6 & 4 \\
$\beta$ Pictoris ($\beta$ Pic) &  12-22 & 13 & 11\\
Octans (Oct)& 20-40 & 1 &1\\
Columba (Col)  & 20-40 & 6 & 4\\
$\eta$ Carinae ($\eta$ Car) &  20-40&  12& 7\\
Tucana-Horologium (Tuc/Hor) & 20-40& 16 & 13\\
Argus (Arg) &  30-50& 10 & 10 \\
IC 2391& 30-50&  3 & 3 \\
AB Doradus (AB Dor) & 70-120&10 & 6\\
\hline
\end{tabular}
\tablefoot{For each association we reported the number of investigated stars and the number of stars where at least one activity cycle was detected. The ages are the same reported in Paper I (see that paper and the references therein for details on the age estimate).}
\end{table*}

\subsection{Computation of the global  parameters for the target stars}
In this paper we investigate the relationships between the cycles lengths $P_{\rm cyc}$, the activity index <IQR>, and the global  parameters of our targets.
Our analysis was focused on the following parameters: the photometric shear $\Delta\Omega_{\rm phot}$,   the effective temperature $T_{\rm eff}$,  the convective turnover time-scale $\tau_{\rm C}$, the Rossby number $Ro$,  the radius at the bottom of the convection zone $R_{\rm C}$, the dynamo number $D_{\rm N}$, and the magnetic diffusivity time-scale $\tau_{\rm diff}$. 
 $\Delta\Omega_{\rm phot}$, $T_{\rm eff}$, $\tau_C$ and $Ro$ have been computed in Paper I where as $R_{\rm C}$, $D_{\rm N}$ and $\tau_{\rm diff}$ have been computed in this work for the first time.

The photometric shear of a given star is defined as 
\begin{equation}
\label{photshear}
\Delta\Omega_{\rm phot}= \Omega_{\rm max} -\Omega_{\rm min} =\frac{2\pi}{P_{\rm min}} -\frac{2\pi}{P_{\rm max}}
\end{equation}
where $P_{\rm min}$ and $P_{\rm max}$ are the minimum and maximum values of the $P_{\rm rot}$ index and are interpreted as the rotation periods of two Active Regions occurring at two different  latitudes. This parameter can be regarded as a lower limit for the SDR that   is defined as
\begin{equation}
\label{photshear}
\Delta\Omega= \Omega_{\rm Eq} -\Omega_{\rm pole} 
\end{equation}
where $\Omega_{\rm Eq}$ and $\Omega_{\rm pole}$ are the stellar rotation frequencies at the equator and at the poles.

The effective temperature $T_{\rm eff}$ and  the convective turnover time-scale $\tau_{\rm C}$ have been derived in Paper I using \cite{2013ApJ...776...87S}.

The Rossby number $Ro$ is defined as:

\begin{equation}
\label{rossbyspada}
Ro\equiv \frac{P_{\rm rot}}{\tau_{\rm C}},
\end{equation}

where $\tau_{\rm C}$ is derived by the theoretical isochrones of \cite{2013ApJ...776...87S}. Other works \citep{1999ApJ...524..295S,2016A&A...588A..38L} use the definition given by \cite{1998ApJ...498L..51B}:

\begin{equation}
\label{rossbybra}
Ro \equiv \frac{P_{\rm rot}}{4\pi \tau_{\rm C}} .
\end{equation}

In these works, the convective turnover time-scale is computed through the  semi-empirical equation given by \cite{1984ApJ...279..763N} that expresses $\tau_{\rm C}$ as a function of the color index $B-V$. In order to make a comparison between our results and those of the above mentioned works, we  computed the  Ro values according to the Eq. (\ref{rossbybra}) and by using  the $\tau_{\rm C}$ values as given by  the formula of \cite{1984ApJ...279..763N}. In the rest of the paper, the values computed by means of Eq. \ref{rossbybra} will be indicated  with the symbol $Ro_{\rm Br}$ where $\rm Br$ stands for Brandenburg.

The dynamo number is defined as:

\begin{equation}
\label{dynamonumber}
D_{\rm N}= \frac{\alpha \Delta\Omega D_{\rm CZ}^3}{\eta_t^2},
\end{equation}
with
\begin{equation}
\label{alphadyn}
\alpha=  \frac{l_{\rm m}^2}{H_D}\Omega,
\end{equation}
where $D_{\rm CZ}$ is the depth of the convective zone, $l_{\rm m}$ is the mixing-length and $H_D$ is the density scale-height, $\eta_t$ is the turbulent diffusivity and $\Delta\Omega$ the rotational shear.
The parameters $D_{\rm CZ}$, $\eta_t$, $l_{\rm m}$ and $H_D$ occurring in Eq. \ref{dynamonumber} and \ref{alphadyn} were derived by the theoretical models of \cite{2013ApJ...776...87S}. The parameter $\Delta\Omega$ was replaced with the parameter $\Delta\Omega_{\rm phot}$ computed in Paper I.

Finally, the magnetic diffusivity time-scale is defined as:
\begin{equation}
\tau_{\rm diff}=\frac{D_{\rm CZ}^2}{\eta_t}
\end{equation}
where $\eta_t$ is the turbulent diffusivity;  $\tau_{\rm diff}$ is the time-scale  for the turbulent dissipation of magnetic energy over a length scale $D_{\rm CZ}$.
All the parameters computed for the different targets are reported in Table \ref{parameters}.

\section{The method}
We searched for activity cycles by running the Lomb-Scargle  \citep{1976Ap&SS..39..447L,1982ApJ...263..835S}  and the PDM \citep[Phase Dispersion Minimization;][]{1971Ap&SS..13..154J,1978ApJ...224..953S} algorithms on the $P_{\rm rot}$, the $V_{\rm med}$ and the IQR time-series. The ASAS time-series  span, on average, an interval T=3000 days. 
 We assumed, as in \cite{2012MNRAS.421.2774D}, that a period P can be detected if $P\le 0.75  T$ and we performed our period search in the range  (100-2000 d).
We computed the FAP (False Alarm Probability) associated with a given period $P_{\rm cyc}$ according to the procedure described in Sec. \ref{fapcomputation} and we flagged a period as valid if it satisfied the requirement $\rm FAP <0.1\%$. 
In some time-series, visual inspection allowed the identification of cyclical patterns that the Lomb-Scargle and the PDM algorithm were not able to detect.
The failure of the period search algorithms is due to the ``quasi-periodic'' nature of the stellar cycles and can be ascribed to different reasons: in some cases a cyclical pattern occurs only in a limited interval of the whole time-series; in other cases a cyclical pattern is visible in the complete  time-series, but the length of the cycle changes in time. 
Finally, in some cases, the visual inspection suggests that the star is characterized by a cycle longer than the whole time-series length. 
Since the size of our sample is relatively small, we decided to visually inspect each time-series  aiming at detecting and  measuring the lengths of the cycles missed by the Lomb-Scargle and the PDM algorithms. In this case, we flagged a $P_{\rm cyc}$ value as valid if at least one complete oscillation is visible (i.e. if at least two consecutive maxima or minima are clearly visible).

The uncertainty on $P_{\rm cyc}$ is estimated through the equation:
\begin{equation}
\delta P_{\rm cyc} = \frac{\delta \nu_{\rm cyc}}{\nu_{\rm cyc}^2}
\end{equation}
where $\delta \nu_{\rm cyc}$ is the uncertainty associated with the frequency of the cycle $\nu_{\rm cyc}=1/P_{\rm cyc}$.  This uncertainty  is given by the equation:
\begin{equation}
\delta \nu_{\rm cyc} = \sqrt{{\delta \nu}_{\rm samp}^2 +\delta\nu_{\rm noise}^2}
\end{equation}
 where $\delta\nu_{\rm samp}$ is the uncertainty due to the limited and discrete sampling of time-series and $\delta\nu_{\rm noise}$ is the uncertainty due to data noise;
 $\delta\nu_{\rm samp}$ and $\delta\nu_{\rm noise}$ are computed according to \citet{1981Apkov81SS..78..175K} through the equations
\begin{equation}
\label{samplingerror}
\delta\nu_{\rm samp}=\frac{0.16}{\sqrt{2} \nu_{\rm cyc}T^2}
.\end{equation}
\begin{equation}
\label{gaussianerror}
\delta\nu_{\rm noise}=\frac{3\sigma}{4\sqrt{N}\nu_{\rm cyc}TA}
\end{equation}
where T is the interval time spanned by the time-series, $\sigma$ is the standard deviation of the data before subtracting any periodic signal, $N$ the number of time-series points, and $A$ the amplitude of the signal.
In cases in which $P_{\rm cyc}$ values were determined by visual inspection  we did not give a  $\delta P_{\rm cyc}$ estimate since there is not an objective way to obtain it.

\subsection{FAP computation}
\label{fapcomputation}
The typical Monte Carlo method used to assess the FAP of a given period $P_0$, that has a power $z_0$ in the Lomb-Scargle periodogram,  consists in simulating N synthetic time-series with the same sampling of the original time-series. The synthetic data-points are usually generated by means  of the equation:
\begin{equation}
\label{UGN}
x_i=<x> + R(0,\sigma)
\end{equation}
where $<x>$ is the mean value of the original time-series and $R(0,\sigma)$ is a gaussian random variate with a zero mean and  a dispersion $\sigma$ given by the standard deviation associated with the real data-points. The Lomb-Scargle periodogram is computed for each time-series and the highest peak $Z_{\rm max}$ is retained for each  periodogram. The FAP associated with $P_0$ is then computed as the fraction of synthetic time-series for which $Z _{\rm max}~\ge z_0$.

The previous approach is valid only if the points of the original time-series are uncorrelated i.e. if two consecutive data points are independent of each other \citep[see for example the discussion in][]{1996AAS...189.4908H,1999AJ....117.2941S,2001AJ....121.1676R}. This is not the case of our time-series where  two consecutive  elements of the activity indexes have been computed in segments that partially overlap with each other.

For this reason, we estimated the FAP by means of the Monte Carlo approach suggested by \cite{1996ApJS..107..263B} and also adopted by \cite{2001AJ....121.1676R} and \cite{2009MNRAS.400..603P}. In this approach, the synthetic data-points are generated by means of a recursive procedure:
\begin{equation}
\begin{array}{l}
\displaystyle x_0= R(0,\sigma) \\
\displaystyle x_{i>0} = \alpha x_{i -1} + \beta R(0,\sigma)
\end{array}
\label{GCN}
\end{equation}
where  $x_0$ and $x_i$  are the first element and the i-th element of the simulated time-series and $R(0, \sigma)$ denotes a Gaussian random variate centered
on zero with dispersion $\sigma$ (equal to the standard deviation of the real data).
The parameters $\alpha$ and $\beta$ are defined as
\begin{equation}
\begin{array}{l}
 \alpha=exp(-\Delta t/L_{corr})\\
 \beta=(1 -\alpha^2)^{1/2}
 \end{array}
 \label{alphabeta}
 \end{equation}
 where $\Delta t = t_i -t_{i-1}$ is the time interval between the two consecutive points $x_{i-1}$ and $x_i$, and $L_{corr}$ is the correlation time-scale. 
The time-series defined by Eqs. (\ref{GCN}) is such that two points $x_i$ and $x_j$ satisfy the property
\begin{equation}
\label{crf}
C_2(i,j)=exp(-\Delta t_{ij}/L_{corr})
\end{equation}
where $C_2(i,j)$ is the two-points correlation function and $\Delta t_{ij}=t_i -t_j$ the  time between $x_i$ and $x_j$ \citep{1996ApJS..107..263B}.
 In our case, we adopted $L_{corr} = T/3$ were T is the length of the sliding window used to generate the activity index time-series.

In order to assess the FAP probability associated with the period detected in a given time-series we followed the Monte Carlo procedure described above by generating 10000 synthetic time-series.
If the period was found by means of the Lomb-Scargle algorithm, the FAP was computed by taking the fraction of synthetic time-series for which $Z_{\rm max} > z_0$.
If the period $P_0$ was found by means of the PDM algorithm,  we run the PDM on each synthetic time-series and we retained the minimum value $\theta_{\rm min}$ of  the PDM periodogram for each time-series.  The FAP was then computed by taking the fraction of the  synthetic time-series for which $\theta_0 < \theta_{\rm min}$, where $\theta_0$ is the PDM estimator associated with $P_0$ .
\begin{figure}
\begin{center}
\includegraphics[width=80mm]{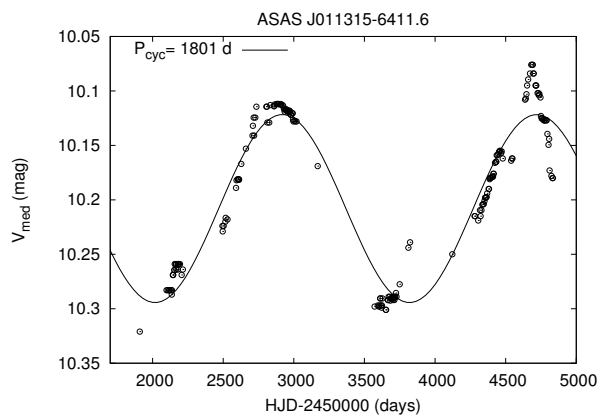}
\caption{The  $V_{\rm med}$ time-series for the star ASAS J011315-6411.6. A cycle with length $P_{\rm cyc}~=~1801~d$ was detected by the period-search algorithms. The black  line is the sinusoid best-fitting the data. }
\label{example0}
\end{center}
\end{figure}

\begin{figure}
\begin{center}
\includegraphics[width=80mm]{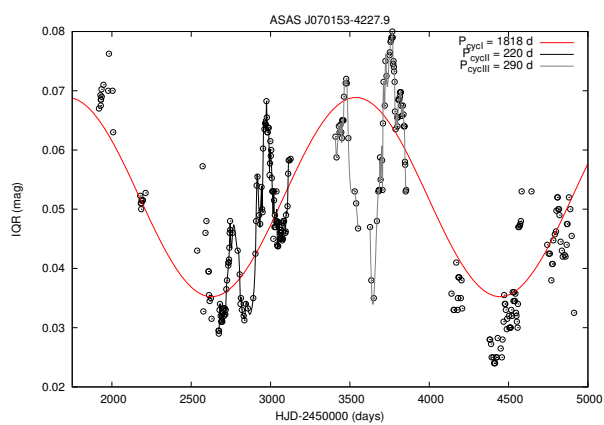}
\caption{The  IQR time-series for the star ASAS J070153-4227.9.  A cycle with length $P_{\rm cyc}~=~1818~d$ (red  line) was detected by the period search algorithms. A visual inspection reveals also two shorter cycles with lengths 220 and 290 d, respectively. The dark and gray lines were obtained by fitting the data with smoothing cubic splines and were plotted to highlight the cycles detected by eye.}
\label{example1}
\end{center}
\end{figure}
 
\begin{figure}
\begin{center}
\includegraphics[width=80mm]{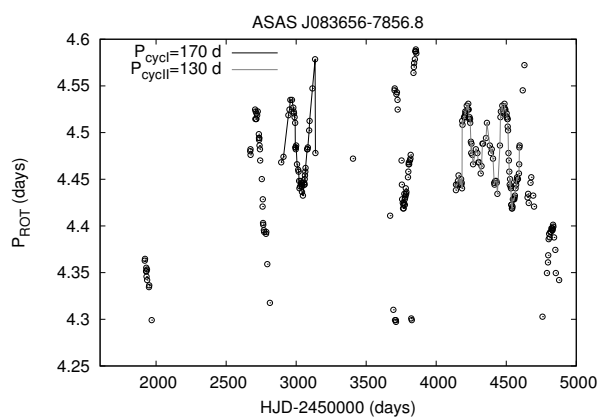}
\caption{The  $P_{\rm rot}$ time-series for the star ASAS J083656-7856.8. A visual inspection revealed two cycles with lengths 170 d and 130 d, respectively. Two smoothing cubic splines are over-plotted to highlight the cycles with the dark and the gray line, respectively. }
\label{example2}
\end{center}
\end{figure}
 
 \begin{figure}
\begin{center}
\includegraphics[width=80mm]{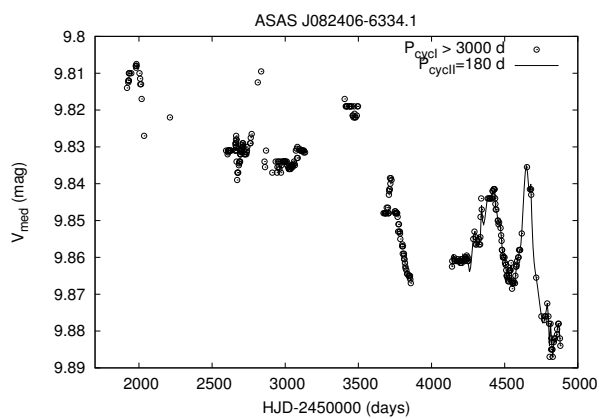}
\caption{The  $V_{\rm med}$ time-series for the star ASAS J082406-6334.1. The time-series shows a long-term trend suggesting a cycle longer than 3000 d. A short cycle of 180 d is also visibile and highlighted with a smoothing cubic spline in black. }
\label{example3}
\end{center}
\end{figure}

\begin{longtab}
\begin{longtable}{llllllllllllll}
\caption{\label{parameters}List of the targets investigated in the present work.}\\
\hline\hline
Target ID & Assoc. & $P_{\rm rot}$ & $\Delta\Omega_{\rm phot}$  & m &$T_{\rm eff}$ &$R_{\rm C}$ & $\tau_{\rm C}$ & $\tau_{\rm diff}$  &$D_{\rm N}$&Ro & $Ro_{\rm Br}$ &<IQR>& \\
               &      &    (d)       &  ($\rm rad~d^{-1}$)        &$M_{\odot}$  & K      &    &  d & d & & &   &       \\
\hline\hline
\endfirsthead
\caption{continued.}\\
\hline\hline
Target ID & Assoc. & $P_{\rm rot}$ & $\Delta\Omega_{\rm phot}$ &  m &$T_{\rm eff}$ &$R_{\rm C}$ & $\tau_{\rm C}$ & $\tau_{\rm diff}$ & $D_{\rm N}$ & Ro & $Ro_{\rm Br}$ &<IQR>& \\
               &      &    (d)       &  ($\rm rad~d^{-1}$)       &$M_{\odot}$  & K      &    &  d & d & & &    &      \\
\hline
\endhead
\hline
\endfoot
   J001353-7441.3 & TUC/HOR & 3.67 & 0.146 & 0.95 & 5060.67 & 0.68 & 50 & 544 & 45182 & 0.074 & 0.017 &0.038\\
  J002409-6211.1 & TUC/HOR & 1.75 & 0.068 & 0.78 & 4325.44 & 0.6 & 146 & 1337 & 192079 & 0.012 & 0.005 &0.052\\
  J003451-6155.0 & TUC/HOR & 0.38 & 0.057 & 1.03 & 5446.5 & 0.74 & 32 & 391 & 98487 & 0.012 & 0.001 &0.051\\
  J004220-7747.7 & TUC/HOR & 2.57 & 0.048 & 0.86 & 4608.72 & 0.63 & 93 & 1013 & 58816 & 0.028 & 0.009 &0.048\\
  J011315-6411.6 & TUC/HOR & 1.26 & 0.045 & 0.83 & 4513.54 & 0.62 & 119 & 1075 & 124198 & 0.011 & 0.005 &0.078\\
  J015749-2154.1 & TUC/HOR & 3.05 & 0.086 & 1.06 & 5596.57 & 0.75 & 24 & 163 & 4490 & 0.124 & 0.02 &0.035\\
  J020136-1610.0 & COL & 3.21 & 0.076 & 1.07 & 5649.15 & 0.75 & 24 & 212 & 5738 & 0.131 & 0.011 &0.055\\
  J020718-5311.9 & TUC/HOR & 2.33 & 0.233 & 1.04 & 5499.37 & 0.74 & 32 & 255 & 32746 & 0.073 & 0.015 &0.031\\
  J024126+0559.3 & BPIC & 4.83 & 0.035 & 0.83 & 4115.02 & 0.55 & 196 & 2036 & 79357 & 0.025 & 0.016 &0.093\\
  J024233-5739.6 & TUC/HOR & 7.4 & 0.071 & 0.77 & 4263.07 & 0.6 & 146 & 1348 & 48069 & 0.051 & 0.024 &0.056\\
  J030942-0934.8 & ABDOR & 5.47 & 0.068 & 0.95 & 5545.63 & 0.72 & 39 & 249 & 3822 & 0.14 & 0.022 &0.028\\
  J031909-3507.0 & TUC/HOR & 8.52 & 0.061 & 0.71 & 4000.66 & 0.57 & 172 & 1701 & 51611 & 0.05 & 0.027 &0.065\\
  J033049-4555.9 & TUC/HOR & 3.8 & 0.122 & 0.89 & 4764.23 & 0.65 & 93 & 833 & 73155 & 0.041 & 0.013 &0.036\\
  J033156-4359.2 & TUC/HOR & 2.93 & 0.07 & 0.74 & 4151.49 & 0.58 & 172 & 1476 & 138328 & 0.017 & 0.009 &0.057\\
  J034723-0158.3 & ABDOR & 3.86 & 0.057 & - & - & - & - & - & - & - & 0.011 &0.070\\
  J045249-1955.0 & TUC/HOR & 5.21 & 0.06 & 1.08 & 5709.79 & 0.76 & 24 & 189 & 2332 & 0.213 & 0.019 &0.062\\
  J045305-4844.6 & COL & 4.59 & 0.059 & 0.9 & 4820.09 & 0.65 & 71 & 755 & 24951 & 0.065 & 0.017 &0.048\\
  J045935+0147.0 & BPIC & 4.42 & 0.047 & 0.8 & 4015.72 & 0.53 & 196 & 2339 & 144854 & 0.023 & 0.013 &0.064\\
  J050047-5715.4 & BPIC & 8.74 & 0.05 & 0.8 & 4005.8 & 0.53 & 196 & 2365 & 79212 & 0.045 & 0.027 &0.049\\
  J050230-3959.2 & ABDOR & 6.58 & 0.073 & 0.68 & 4320.15 & 0.68 & 75 & 752 & 18279 & 0.088 & 0.024 &0.074\\
  J050651-7221.2 & Octans & 0.24 & 0.101 & - & - & - & - & - & - & - & - &0.044\\
  J052845-6526.9 & ABDOR & 0.51 & 0.053 & 0.93 & 5455.8 & 0.71 & 45 & 294 & 41297 & 0.011 & 0.002 &0.068\\
  J052857-3328.3 & ABDOR & 0.69 & 0.07 & 0.88 & 5274.9 & 0.7 & 50 & 365 & 56018 & 0.014 & 0.002 &0.045\\
  J053705-3932.4 & TUC/HOR & 2.46 & 0.182 & 0.95 & 5038.04 & 0.68 & 50 & 597 & 97627 & 0.05 & 0.01 &0.037\\
  J055101-5238.2 & COL & 1.2 & 0.081 & 1.02 & 5392.94 & 0.73 & 32 & 343 & 35666 & 0.038 & 0.006 &0.047\\
  J055329-8156.9 & CAR & 1.86 & 0.125 & 1.09 & 5760.17 & 0.76 & 24 & 87 & 4062 & 0.076 & 0.008 &0.050\\
  J055751-3804.1 & ABDOR & 0.79 & 0.143 & 1.04 & 5862.33 & 0.74 & 32 & 153 & 26084 & 0.024 & 0.004 &0.041\\
  J060834-3402.9 & ABDOR & 3.4 & 0.1 & - & - & - & - & - & - & - & 0.014 &0.052\\
  J061828-7202.7 & BPIC & 2.67 & 0.045 & 0.88 & 4302.06 & 0.57 & 173 & 1504 & 113230 & 0.015 & 0.009 &0.052\\
  J062607-4102.9 & COL & 4.18 & 0.106 & 1.07 & 5623.8 & 0.75 & 24 & 176 & 4554 & 0.171 & 0.017 &0.048\\
  J062806-4826.9 & COL & 1.29 & 0.078 & 1.04 & 5490.63 & 0.74 & 32 & 296 & 25239 & 0.041 & 0.009 &0.049\\
  J063950-6128.7 & ABDOR & 9.12 & 0.045 & 0.69 & 4383.69 & 0.68 & 75 & 750 & 8104 & 0.121 & 0.029 &0.027\\
  J064346-7158.6 & CAR & 3.89 & 0.128 & 1.02 & 5370.69 & 0.73 & 32 & 324 & 15893 & 0.122 & 0.021 &0.047\\
  J065623-4646.9 & COL & 4.45 & 0.124 & 1.02 & 5390.86 & 0.73 & 32 & 376 & 17174 & 0.14 & 0.019 &0.052\\
  J070030-7941.8 & CAR & 5.11 & 0.08 & 0.96 & 5097.57 & 0.69 & 50 & 563 & 18785 & 0.103 & 0.018 &0.060\\
  J070153-4227.9 & Argus & 3.99 & 0.114 & - & - & - & - & - & - & - & 0.015 &0.049\\
  J072124-5720.6 & CAR & 4.67 & 0.114 & 0.97 & 5115.58 & 0.7 & 50 & 550 & 28233 & 0.094 & 0.033 &0.056\\
  J072822-4908.6 & Argus & 1.03 & 0.161 & 1.02 & 5578.2 & 0.74 & 32 & 157 & 23560 & 0.032 & 0.004 &0.043\\
  J072851-3014.8 & ABDOR & 1.64 & 0.047 & 0.68 & 4336.36 & 0.67 & 75 & 761 & 47861 & 0.022 & 0.005 &0.044\\
  J073547-3212.2 & Argus & 5.06 & 0.087 & 0.97 & 5390.66 & 0.73 & 35 & 201 & 3796 & 0.143 & 0.023 &0.025\\
  J082406-6334.1 & CAR & 0.79 & 0.188 & 1.11 & 5886.94 & 0.77 & 22 & 44 & 5001 & 0.037 & 0.006 &0.034\\
  J082844-5205.7 & IC2391 & 1.51 & 0.094 & 1.03 & 5615.66 & 0.74 & 32 & 136 & 7526 & 0.047 & 0.008 &0.057\\
  J083656-7856.8 & $\eta$  Cha & 4.45 & 0.07 & - & - & - & - & - & - & - & 0.014 &0.068\\
  J084006-5338.1 & IC2391 & 1.34 & 0.085 & 1.1 & 5945.58 & 0.76 & 21 & 69 & 2707 & 0.064 & 0.014 &0.038\\
  J084200-6218.4 & CAR & 1.22 & 0.097 & 1.11 & 5858.1 & 0.77 & 22 & 57 & 2509 & 0.057 & 0.005 &0.077\\
  J084229-7903.9 & $\eta$  Cha & 7.14 & 0.087 & 0.82 & 3974.45 & 0.42 & 307 & 2783 & 229195 & 0.023 & 0.022 &0.151\\
  J084300-5354.1 & IC2391 & 3.15 & 0.072 & 1.05 & 5709.71 & 0.75 & 27 & 130 & 2579 & 0.116 & 0.018 &0.069\\
  J084432-7846.6 & $\eta$  Cha & 20.26 & 0.07 & 0.75 & 3870.19 & 0.38 & 339 & 2991 & 69787 & 0.06 & 0.063 &0.108\\
  J084708-7859.6 & $\eta$  Cha & 4.84 & 0.111 & 1.12 & 4690.61 & 0.55 & 176 & 1506 & 182630 & 0.028 & 0.017 &0.161\\
  J085005-7554.6 & CAR & 1.15 & 0.117 & 0.96 & 5073.41 & 0.69 & 50 & 570 & 124376 & 0.023 & 0.005 &0.044\\
  J085156-5355.9 & CAR & 1.91 & 0.226 & 1.11 & 5843.03 & 0.77 & 22 & 59 & 3929 & 0.089 & 0.011 &0.045\\
  J085746-5408.6 & CAR & 1.94 & 0.084 & 1.04 & 5475.28 & 0.74 & 32 & 307 & 19150 & 0.061 & 0.007 &0.080\\
  J085752-4941.8 & CAR & 2.03 & 0.143 & 1.03 & 5456.74 & 0.74 & 32 & 275 & 26042 & 0.064 & 0.01 &0.043\\
  J085929-5446.8 & CAR & 0.44 & 0.111 & 1.08 & 5683.11 & 0.76 & 24 & 80 & 13525 & 0.018 & 0.004 &0.046\\
  J092335-6111.6 & CAR & 3.89 & 0.066 & 1.04 & 5496.23 & 0.74 & 32 & 318 & 7956 & 0.122 & 0.015 &0.054\\
  J092854-4101.3 & Argus & 0.39 & 0.065 & 1.06 & 5751.23 & 0.75 & 27 & 136 & 20150 & 0.014 & 0.002 &0.069\\
  J094247-7239.8 & Argus & 2.31 & 0.07 & 0.9 & 5090.13 & 0.7 & 41 & 554 & 33175 & 0.056 & 0.015 &0.104\\
  J095558-6721.4 & Argus & 1.83 & 0.085 & 1.13 & 6066.08 & 0.78 & 22 & 48 & 1104 & 0.083 & 0.016 &0.032\\
  J101315-5230.9 & TWA & 4.4 & 0.094 & 0.9 & 4147.98 & 0.5 & 219 & 2366 & 327097 & 0.02 & 0.015 &0.050\\
  J105351-7002.3 & Argus & 1.03 & 0.102 & 1.14 & 6140.48 & 0.78 & 22 & 31 & 1226 & 0.047 & 0.008 &0.045\\
  J105749-6914.0 & $\epsilon$  Cha & 3.58 & 0.053 & 1.2 & 4685.66 & 0.46 & 234 & 1704 & 159865 & 0.015 & 0.013 &0.077\\
  J110914-3001.7 & TWA & 4.85 & 0.086 & 0.9 & 4146.39 & 0.5 & 219 & 2427 & 275253 & 0.022 & 0.014 &0.067\\
  J112105-3845.3 & TWA & 3.3 & 0.039 & - & - & - & - & - & - & - & 0.01 &0.186\\
  J112117-3446.8 & TWA & 5.44 & 0.079 & 0.84 & 4013.62 & 0.47 & 269 & 2790 & 273721 & 0.02 & 0.016 &0.094\\
  J112205-2446.7 & TWA & 14.3 & 0.056 & - & - & - & - & - & - & - & 0.047 &0.033\\
  J115942-7601.4 & $\epsilon$  Cha & 8.04 & 0.052 & 0.86 & 4131.34 & 0.31 & 399 & 2525 & 105940 & 0.02 & 0.027 &0.074\\
  J120139-7859.3 & $\epsilon$  Cha & 4.38 & 0.124 & - & - & - & - & - & - & - & 0.032 &0.030\\
  J120204-7853.1 & $\epsilon$  Cha & 4.44 & 0.041 & 0.88 & 4162.21 & 0.32 & 399 & 2550 & 152747 & 0.011 & 0.015 &0.202\\
  J121138-7110.6 & $\epsilon$  Cha & 5.13 & 0.098 & - & - & - & - & - & - & - & 0.025 &0.040\\
  J121531-3948.7 & TWA & 5.06 & 0.047 & 0.87 & 4076.6 & 0.49 & 244 & 2562 & 155692 & 0.021 & 0.015 &0.125\\
  J122023-7407.7 & $\epsilon$  Cha & 1.54 & 0.077 & 0.83 & 4073.53 & 0.29 & 443 & 2605 & 844548 & 0.003 & 0.005 &0.158\\
  J122034-7539.5 & Argus & 3.49 & 0.088 & 0.75 & 4436.18 & 0.62 & 103 & 1046 & 78688 & 0.034 & 0.012 &0.051\\
  J122105-7116.9 & $\epsilon$  Cha & 6.86 & 0.05 & 0.82 & 4052.4 & 0.29 & 443 & 2667 & 126155 & 0.015 & 0.022 &0.091\\
  J123921-7502.7 & $\epsilon$  Cha & 3.99 & 0.05 & 1.09 & 4514.39 & 0.42 & 283 & 1890 & 151791 & 0.014 & 0.014 &0.054\\
  J125826-7028.8 & $\epsilon$  Cha & 2.0 & 0.048 & 1.23 & 4715.63 & 0.47 & 234 & 1673 & 253207 & 0.009 & 0.008 &0.043\\
  J134913-7549.8 & Argus & 2.29 & 0.1 & 1.06 & 5745.5 & 0.75 & 27 & 118 & 4226 & 0.084 & 0.013 &0.039\\
  J153857-5742.5 & BPIC & 4.3 & 0.105 & 1.14 & 5642.4 & 0.68 & 68 & 339 & 14521 & 0.063 & 0.017 &0.043\\
  J171726-6657.1 & BPIC & 1.68 & 0.119 & - & - & - & - & - & - & - & 0.008 &0.033\\
  J181411-3247.5 & BPIC & 2.42 & 0.097 & 1.09 & 5366.58 & 0.65 & 87 & 601 & 61130 & 0.028 & 0.009 &0.039\\
  J181952-2916.5 & BPIC & 0.57 & 0.097 & - & - & - & - & - & - & - & 0.003 &0.056\\
  J184653-6210.6 & BPIC & 5.37 & 0.053 & 0.78 & 3955.37 & 0.52 & 222 & 2583 & 156477 & 0.024 & 0.016 &0.131\\
  J185306-5010.8 & BPIC & 0.94 & 0.072 & 1.14 & 5651.12 & 0.68 & 68 & 327 & 43013 & 0.014 & 0.004 &0.036\\
  J200724-5147.5 & Argus & 0.84 & 0.036 & 0.62 & 3826.1 & 0.56 & 185 & 2097 & 405326 & 0.005 & 0.003 &0.054\\
  J204510-3120.4 & BPIC & 4.84 & 0.065 & 0.62 & 3721.1 & 0.42 & 314 & 3531 & 318887 & 0.015 & 0.014 &0.053\\
  J205603-1710.9 & BPIC & 3.4 & 0.042 & 0.91 & 4447.94 & 0.58 & 147 & 1360 & 70789 & 0.023 & 0.011 &0.050\\
  J212050-5302.0 & TUC/HOR & 3.43 & 0.195 & 0.99 & 5233.2 & 0.71 & 50 & 432 & 43925 & 0.069 & 0.017 &0.039\\
  J214430-6058.6 & TUC/HOR & 4.55 & 0.086 & 0.65 & 3823.77 & 0.53 & 200 & 2464 & 245430 & 0.023 & 0.014 &0.065\\
  J232749-8613.3 & TUC/HOR & 0.7 & 0.168 & 0.99 & 5252.98 & 0.71 & 50 & 415 & 173598 & 0.014 & 0.006 &0.050\\
  J233231-1215.9 & BPIC & 5.69 & 0.031 & 0.73 & 3859.11 & 0.49 & 254 & 3050 & 110692 & 0.022 & 0.017 &0.055\\
  J234154-3558.7 & ABDOR & 1.79 & 0.147 & 0.99 & 5696.24 & 0.73 & 39 & 197 & 17603 & 0.046 & 0.009 &0.044\\  
\end{longtable}
\end{longtab}
\section{Results}
We  detected activity cycles in 67 stars.We found two distinct cycles in 32 of them and  more than two cycles in 16 stars. 
The results of our analysis are reported in Table \ref{results}.
In this table we listed the $P_{\rm cyc}$ lengths found for each star. For each cycle we reported  a flag indicating the activity index from  which $P_{\rm cyc}$ was inferred and a second flag indicating the method used to detect the period. If $P_{\rm cyc}$ was inferred by means of the Lomb-Scargle or the PDM algorithm, we also reported the associated FAP. 

In Figs. 1-4  we report some  examples of our analysis. 
In Fig. \ref{example0} we display the $V_{\rm med}$ time-series for the star ASAS J011315-6411.6 .
The period search algorithms detected in this case a cycle with $P_{\rm cyc}~=~1801~d$. The sinusoid best-fitting the data is over-plotted (red continuous line).

In Fig. \ref{example1}  we plot the  $IQR$ time-series for the  star ASAS J070153-4227.9. The 
Lomb-Scargle periodgram of this time-series gives a highly significant peak at P=1818 d. However, visual inspection  reveals also two shortest cycles with lengths of 220 and 290 d, respectively. 
These  cycles could be secondary cycles analogue to the Rieger cycles or to the quasi-biennal oscillations over-imposed on the 11-yr solar cycle.

In Fig.\ref{example2} we report the  $P_{\rm rot}$ time-series for the star ASAS 083656-7856.8. The Lomb-Scargle and PDM algorithms were not able to detect any significant period in this star. However two cycles with lengths 170 d and 130 d are clearly visible in the data.
These features are very common in the Sun and in solar-like stars.
For instance, the Rieger cycles and the quasi-biennal oscillations have been well detected in the Sun but, as remarked by \cite{2016A&A...590A.133O}, they are not continuously present in the solar data registered until now.
\cite{2016A&A...590A.133O} noticed that in the Mount Wilson time-series some of the detected cycles are only temporarily seen and that young stars are characterized by cycles whose duration changes in time. Rieger-like cycles were also detected by \cite{2009A&A...493..193L} in CoRoT-2 and by \cite{2012A&A...547A..37B} in Kepler-17.

Finally in Fig. \ref{example3}, we report the $V_{\rm med}$ time-series of the star ASAS J082406-6334.1. The visual inspection suggests the existence of a cycle longer than 3000 d. A shorter cycle with $P_{\rm cyc}~=180~\rm d$ is also detected by eye and highlighted with the black continuous line.
Note that some of the $P_{\rm cyc}$ values detected by visual inspection are shorter than the sliding-window length used to process the ASAS time-series. In fact, the 100-d sliding window attenuate the periodic signals shorter than 100-d but, in some cases, does not completely suppress them. So if these signal have a sufficient amplitude, they can still be detected after the segmentation procedure(see Appendix \ref{signals} for details.

\subsection{The case of AB Dor A}
To the best of our knowledge, the cycles reported in Table \ref{results} have been detected here for the first time.
The only exception in our sample is given by   AB Dor A (HD 36705)  that is identified with the ASAS ID J052845-6526.9.
AB Dor A is a fast rotating ($P_{\rm rot }=0.51 d$) K1V star well studied in the literature and is the primary component of the quadruple system AB Dor that comprises also the stars AB Dor Ba, AB Dor Bb and AB Dor C.
AB Dor Ba and AB Dor Bb are the components of the binary system AB Dor B that was resolved for the first time by \cite{2007A&A...462..615J} and is located  at $8.9''\pm 0.1''$  from AB Dor A \citep{1995A&A...294..744M}. 
AB Dor C is a close companion of AB Dor A. It was detected by \cite{1997ApJ...490..835G} and it is located at about 0.16''  from AB Dor A.
\cite{2010ASSP...14..139G} made a dynamical estimate of  the AB Dor A mass  and obtained $M=0.86\pm0.09 M_{\odot}$ that is in good agreement with the photometric estimate $M=0.93 M_{\odot}$ made in Paper I and based on the theoretical isochrones of \cite{2013ApJ...776...87S} (see Paper I for details).

 The magnetic activity of AB Dor A has been widely studied in the literature by means of spectroscopic and photometric data collected at different wavelengths  \citep[see e.g.][and references therein]{2015ApJ...802...62D,2013A&A...559A.119L,2009AN....330..358B,2013A&A...559A.119L,2007MNRAS.375..567J,2006A&A...447..293M,2005A&A...432..657J}

\begin{figure}
\begin{center}
\includegraphics[width=80mm]{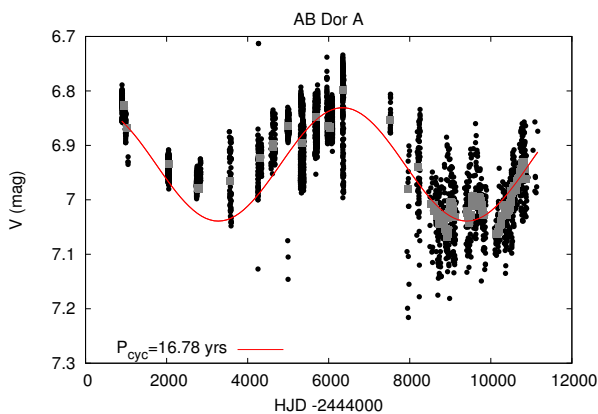}
\caption{The  $V_{\rm med}$ time-series for the AB Dor A. The grey squares are used to mark the $V_{\rm med}$ values computed in the 50-days segments. }
\label{abdor}
\end{center}
\end{figure}

 \cite{2005A&A...432..657J} merged the photometric data collected in several works and obtained a  V-band time-serie spanning the years 1978-2000. They analyzed these data with the inversion technique developed by \cite{1998A&A...338...97B} in order to study the temporal evolution of the spots longitude distribution.  
They  analyzed the temporal variations of the spots longitudes and of the  mean brightness by means of a Fourier analysis  and detected  a primary cycle with length $P_{\rm cyc1}= 21 \pm 3 ~\rm yr$ and a secondary cycle with length $P_{\rm cyc2}\sim5.5 ~\rm yr$. The primary cycle is mainly associated with the mean magnitude variations while the secondary one is a "flip-flop'' cycle i.e. a periodic switch of the longitude where the dominant spot concentration occurs.
 \cite{2005A&A...432..657J} remarked that their estimate of the primary cycle length is not very accurate because of the sparseness of data in the years 1978-1985. 
We merged the photometric data reported in literature  and summarized in   \cite{2005A&A...432..657J} with the ASAS data and we obtained a 27-years V-band  time-series. We processed this long-term data-set with the sliding window algorithm described in Paper I and we obtained three 27-years time-series for the activity indexes $V_{\rm med}$, $P_{\rm rot}$ and IQR. 
Note that, in this specific case, we used a 50-d sliding-window because the duration of the different observing seasons usually  do not exceed two months and so, the use of a 100-d sliding window should be meaningless. 
The  analysis of the $V_{\rm med}$ time-series with the Lomb-Scargle algorithm revealed   an activity cycle of length $P_{\rm cyc}= 16.78  \pm 2 ~\rm yr $.
In Fig. \ref{abdor} we plot the long-term photometric time-series and the sinusoid best-fitting the data. 

The use of the activity indexes  time-series allowed also the detection of secondary cycles with lengths of 400 d, 190 d and 90 d, respectively.

\begin{longtab}
\begin{longtable}{lllllll}
\caption{\label{results}List of the cycles detected in the present work}\\
\hline\hline
Target ID & $P_{\rm cyc}$ & Err & FAP & Meth. & Act. Ind &Mult.   \\
               &         (days)       &  (days)            &\%  &       &  &           \\
\hline\hline
\endfirsthead
\caption{continued.}\\
\hline\hline
Target ID & $P_{\rm cyc}$ & Err & FAP & Meth. & Act. Ind &Mult.   \\
               &         (days)       &  (days)            & \% &       &  &           \\
\hline
\endhead
\hline
\endfoot
 ASAS J001353-7441.3 & 1600 & 212 & < 0.01 & PDM & M & I\\ 
ASAS J002409-6211.1 & 1025 & 52 & 0.07 & PDM & AM & I\\ 
ASAS J002409-6211.1 & 390 & & - & V & M & II\\ 
ASAS J002409-6211.1 & 190 &  & - & V & A & III\\ 
ASAS J002409-6211.1 & 100 & - & - & V & P & IV\\ 
ASAS J003451-6155.0 & >3000 & - & - & V & M & I\\ 
ASAS J003451-6155.0 & 360 & - & - & V & M & II\\ 
ASAS J003451-6155.0 & 290 & - & - & V & A & III\\ 
ASAS J011315-6411.6 & 1801 & 161 & < 0.01 & LS & M & I\\ 
ASAS J011315-6411.6 & 1428 & 172 & < 0.01 & PDM & A & II\\ 
ASAS J011315-6411.6 & 290 & - & - & V & A & III\\ 
ASAS J015749-2154.1 & >3000 & -- & -- & V & M & I\\ 
ASAS J020136-1610.0 & 615 & 35 & 0.08 & PDM & A & I\\ 
ASAS J020136-1610.0 & 336 & 9 & 0.1 & LS & M & II\\ 
ASAS J020718-5311.9 & >3000 & -- & -- & V & M & I\\ 
ASAS J020718-5311.9 & 1052 & 812 & < 0.01 & PDM & P & II\\ 
ASAS J024126+0559.3 & 1249 & 362 & 0.04 & PDM & AM & I\\ 
ASAS J024233-5739.6 & 190 & - & - & V & A & II\\ 
ASAS J033049-4555.9 & 1081 & 80 & < 0.01 & PDM & A & I\\ 
ASAS J033156-4359.2 & >3000 & -- & -- & V & A & I\\ 
ASAS J033156-4359.2 & 930 & 42 & 0.09 & PDM & M & II\\ 
ASAS J045935+0147.0 & 1597 & 300 & < 0.01 & PDM & AM & I\\ 
ASAS J050047-5715.4 & 1666 & 192 & < 0.01 & PDM & M & I\\ 
ASAS J050047-5715.4 & 1142 & 134 & < 0.01 & PDM & A & II\\ 
ASAS J050047-5715.4 & 380 & - & - & V & M & III\\ 
ASAS J050651-7221.2 & 1081 & 80 & < 0.01 & PDM & AM & I\\ 
ASAS J052845-6526.9 & >3000 & - & - & V & M & I\\ 
ASAS J052845-6526.9 & 400 & - & - & V & A & II\\ 
ASAS J052845-6526.9 & 190 & - & - & V & MP & III\\ 
ASAS J052845-6526.9 & 90 & - & - & V & A & IV\\ 
ASAS J053705-3932.4 & 1052 & 50 & 0.01 & PDM & M & I\\ 
ASAS J053705-3932.4 & 370 & - & - & V & A & II\\ 
ASAS J053705-3932.4 & 290 & - & - & V & M & III\\ 
ASAS J055329-8156.9 & 1290 & 56 & 0.04 & PDM & M & I\\ 
ASAS J055329-8156.9 & 530 & - & - & V & A & II\\ 
ASAS J055329-8156.9 & 400 & - & - & V & M & III\\ 
ASAS J055329-8156.9 & 330 & - & - & V & AM & IV\\ 
ASAS J055329-8156.9 & 135 & - & - & V & P & V\\ 
ASAS J055751-3804.1 & 1379 & 102 & 0.03 & PDM & M & I\\ 
ASAS J055751-3804.1 & 95 & - & - & V & P & II\\ 
ASAS J060834-3402.9 & 1333 & 202 & < 0.01 & PDM & AM & I\\ 
ASAS J061828-7202.7 & >3000 & -- & -- & V & M & I\\ 
ASAS J062607-4102.9 & 1538 & 138 & 0.07 & PDM & M & I\\ 
ASAS J062806-4826.9 & 1250 & 117 & 0.02 & PDM & AM & I\\ 
ASAS J063950-6128.7 & 429 & 13 & 0.09 & LS & A & I\\ 
ASAS J064346-7158.6 & >3000 & - & - & V & M & I\\ 
ASAS J064346-7158.6 & 95 & - & - & V & P & II\\ 
ASAS J070030-7941.8 & 1835 & 205 & 0.05& LS & M & I\\ 
ASAS J070030-7941.8 & 250 & - & - & V & AMP & II\\ 
ASAS J070030-7941.8 & 150 & - & - & V & P & III\\ 
ASAS J070153-4227.9 & 1835 & 173 & < 0.01 & LS & A & I\\ 
ASAS J070153-4227.9 & 606 & 48 & < 0.01 & PDM & P & II\\ 
ASAS J070153-4227.9 & 290 & - & - & V & A & III\\ 
ASAS J070153-4227.9 & 230 & - & - & V & A & IV\\ 
ASAS J072124-5720.6 & 1156 & 78 & < 0.01 & LS & AM & I\\ 
ASAS J072124-5720.6 & 190 & - & - & V & A & II\\ 
ASAS J072124-5720.6 & 160 & - & - & V & P & III\\ 
ASAS J072822-4908.6 & 1600 & 157 & 0.06 & PDM & A & I\\ 
ASAS J072851-3014.8 & 1904 & 275 & < 0.01 & PDM & A & I\\ 
ASAS J072851-3014.8 & 1379 & 105 & 0.03 & PDM & M & II\\ 
ASAS J073547-3212.2 & 1069 & 65 & < 0.01 & LS & M & I\\ 
ASAS J082406-6334.1 & >3000 & -- & - & V & M & I\\ 
ASAS J082406-6334.1 & 290 & - & - & V & M & II\\ 
ASAS J082406-6334.1 & 240 & - & - & V & AM & III\\ 
ASAS J082844-5205.7 & 1905 & 226 & 0.03 & PDM & A & I\\ 
ASAS J083656-7856.8 & >3000 & -- & - & V & M & I\\ 
ASAS J083656-7856.8 & 280 & - & - & V & M & II\\ 
ASAS J083656-7856.8 & 235 & - & - & V & A & III\\ 
ASAS J083656-7856.8 & 170 & - & - & V & AP & IV\\ 
ASAS J083656-7856.8 & 130 & - & - & V & P & V\\ 
ASAS J084006-5338.1 & 1905 & 194 & 0.08 & PDM & M & I\\ 
ASAS J084200-6218.4 & 1333 & 123 & < 0.01 & PDM & M & I\\ 
ASAS J084200-6218.4 & 90 & - & - & V & P & II\\ 
ASAS J084229-7903.9 & >3000 & - & - & V & M & I\\ 
ASAS J084229-7903.9 & 420 & - & - & V & M & II\\ 
ASAS J084229-7903.9 & 220 & - & - & V & A & III\\ 
ASAS J084300-5354.1 & >3000 & - & - & V & M & I\\ 
ASAS J084300-5354.1 & 360 & - & - & V & M & II\\ 
ASAS J084300-5354.1 & 200 & - & - & V & M & III\\ 
ASAS J084432-7846.6 & 1481 & 96 & < 0.01 & PDM & M & I\\ 
ASAS J084432-7846.6 & 220 & - & - & V & A & II\\ 
ASAS J084432-7846.6 & 170 & - & - & V & A & III\\ 
ASAS J084708-7859.6 & 400 & - & - & V & M & I\\ 
ASAS J084708-7859.6 & 305 & - & - & V & AM & II\\ 
ASAS J085156-5355.9 & 1587 & 143 & < 0.01 & LS & M & I\\ 
ASAS J085156-5355.9 & 320 & - & - & V & A & II\\ 
ASAS J085929-5446.8 & 590 & 27 & 0.1 & LS & A & I\\ 
ASAS J085929-5446.8 & 438 & 10 & < 0.01 & LS & P & II\\ 
ASAS J085929-5446.8 & 290 & - & - & V & M & III\\ 
ASAS J085929-5446.8 & 150 & - & - & V & A & IV\\ 
ASAS J092854-4101.3 & 1600 & 164 & 0.02 & PDM & M & I\\ 
ASAS J094247-7239.8 & 1156 & 80 & 0.04 & LS & M & I\\ 
ASAS J095558-6721.4 & 1333 & 114 & < 0.01 & PDM & M & I\\ 
ASAS J095558-6721.4 & 365 & - & - & V & M & II\\ 
ASAS J105351-7002.3 & 1599 & 426 & < 0.01 & PDM & M & I\\ 
ASAS J105749-6914.0 & 1639 & 160 & 0.02 & LS & M & I\\ 
ASAS J105749-6914.0 & 394 & 9 & < 0.01 & V & A & II\\ 
ASAS J110914-3001.7 & 1489 & 110 & < 0.01 & PDM & AM & I\\ 
ASAS J112117-3446.8 & 1250 & 138 & < 0.01 & PDM & AM & I\\ 
ASAS J112117-3446.8 & 70 & - & - & V & A & II\\ 
ASAS J112205-2446.7 & 1600 & 196 & 0.01 & PDM & AM & I\\ 
ASAS J115942-7601.4 & 1250 & 112 & < 0.01 & PDM & A & I\\ 
ASAS J115942-7601.4 & 697 & 32 & 0.08 & LS & M & II\\ 
ASAS J115942-7601.4 & 75 & - & - & V & A & III\\ 
ASAS J121531-3948.7 & 1666 & 164 & < 0.01 & PDM & A & I\\ 
ASAS J121531-3948.7 & 1250 & 84 & 0.06 & PDM & M & II\\ 
ASAS J121531-3948.7 & 55 & - & - & V & P & III\\ 
ASAS J122034-7539.5 & 913 & 37 & < 0.01 & LS & A & I\\ 
ASAS J122105-7116.9 & 2000 & 350 & < 0.01 & PDM & A & I\\ 
ASAS J122105-7116.9 & 400 & 7 & < 0.01 & LS & M & II\\ 
ASAS J123921-7502.7 & 1219 & 77 & 0.02 & LS & A & I\\ 
ASAS J123921-7502.7 & 365 & - & - & V & M & II\\ 
ASAS J125826-7028.8 & 1198 & 81 & < 0.01 & LS & M & I\\ 
ASAS J125826-7028.8 & 833 & 36 & 0.07 & PDM & A & II\\ 
ASAS J125826-7028.8 & 80 & - & - & V & A & III\\ 
ASAS J134913-7549.8 & 1999 & 240 & < 0.01 & PDM & M & I\\ 
ASAS J171726-6657.1 & 1904 & 205 & 0.08 & PDM & A & I\\ 
ASAS J181952-2916.5 & 615 & 43 & < 0.01 & PDM & A & I\\ 
ASAS J181952-2916.5 & 90 & - & - & V & P & II\\ 
ASAS J184653-6210.6 & 1999 & 712 & < 0.01 & PDM & A & I\\ 
ASAS J185306-5010.8 & 1136 & 87 & 0.07 & LS & AM & I\\ 
ASAS J200724-5147.5 & 930 & 91 & < 0.01 & PDM & AM & I\\ 
ASAS J204510-3120.4 & 1428 & 283 & < 0.01 & PDM & M & I\\ 
ASAS J204510-3120.4 & 1176 & 171 & 0.01 & PDM & AM & II\\ 
ASAS J205603-1710.9 & 1176 & 220 & < 0.01 & PDM & AM & I\\ 
ASAS J205603-1710.9 & 394 & 11 & 0.03 & LS & M & II\\ 
ASAS J212050-5302.0 & 1315 & 118 & 0.1 & LS & M & I\\ 
ASAS J214430-6058.6 & 1600 & 166 & 0.04 & PDM & A & I\\ 
ASAS J232749-8613.3 & 629 & 24 & 0.1 & LS & M & I\\ 
ASAS J232749-8613.3 & 300 & - & - & V & M & II\\ 
ASAS J232749-8613.3 & 265 & - & - & V & A & III\\ 
ASAS J232749-8613.3 & 165 & - & - & V & A & IV\\ 
ASAS J232749-8613.3 & 85 & - & - & V & AP & V\\ 
ASAS J233231-1215.9 & 1695 & 259 & 0.1 & LS & M & I\\ 
ASAS J234154-3558.7 & 1739 & 268 & 0.1 & PDM & A & I\\ 

\end{longtable}
\tablefoot{Target ID: Asas ID of the investigated source. $P_{\rm cyc}$: cycle length. Err: the error associated with $P_{\rm cyc}$. FAP: False Alarm Probability associated with $P_{\rm cyc}$ (it is given only if $P_{\rm cyc}$ is computed through Lombe-Scargle or PDM). Meth.: a flag indicating the method used to detect the cycle: LS stands for Lomb-Scargle, PDM for Phase-Dispersion-Minimization and v stands for visual inspection. Act. Ind: the activity index time-series from which the cycle was detected. A stands for Amplitude, M for Median Magnitude and P for the rotation period. Mult.: a flag indicating  the sequential number of the detected cycles. The cycles detected in a given star were ranked from the longest to the shortest. }
\end{longtab}
~
~
\section{Discussion}

\subsection{Relationships between the cycle length,  the rotation period and the Rossby number }
Since stellar activity cycles were first detected, a relationship has been searched  between the cycle period $P_{\rm cyc}$ and the stellar rotation period $P_{\rm rot}$.
\cite{2007ApJ...657..486B} analyzed a sample of stars obtained by selecting the best quality spectrophotometric data collected at Mt. Wilson. She identified three different stellar sequences that have an almost constant ratio $\frac{P_{\rm cyc}}{P_{\rm rot}}$.
The Aa and the Ab sequences (where A stands for active) consist of young and active stars. The ratio $\frac{P_{\rm cyc}}{P_{\rm rot}}$ is between 400 and 500 for the Aa stars and it is about 300 for the Ab stars.
The I sequence (where I stands for inactive) comprises old and less active stars for which the ratio $\frac{P_{\rm cyc}}{P_{\rm rot}}$ is about 90.
\cite{2007ApJ...657..486B} speculates that the three sequences are due to different kind of dynamos excited by different phenomena. According to her interpretation, the dynamo should be generated by SDR in A sequence stars and by a vertical shear in I sequence stars. 

\begin{figure}
\begin{center}
\includegraphics[width=80mm]{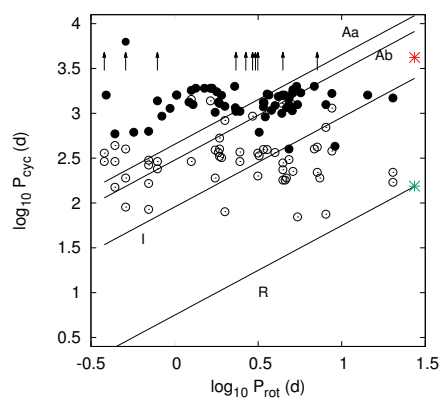}
\caption{The $P_{\rm cyc}$ length vs. the stellar rotation period $P_{\rm rot}$. The filled circles mark the primary cycles  and the empty circles the secondary ones. The arrows mark the stars for which the visual inspection of time-series revealed  $P_{\rm cyc} \ge 3000 d$. The continuous line, the dashed line and the dotted line represent the Aa, the Ab and the I  sequences   identified by \protect\cite{2007ApJ...657..486B}, respectively. 
The dash-dotted line represents the locus where the $\frac{P_{\rm cyc}}{P_{\rm rot}}$ ratio is equal to the ratio between the typical length of the Rieger cycle and the solar rotation period. Red   asterisk: solar Schwabe cycle. Green asterisk: solar Rieger cycle }
\label{pcycprot}
\end{center}
\end{figure}

In Fig. \ref{pcycprot} we plot the cycle lengths vs. the  stellar rotation periods as in \cite{2007ApJ...657..486B}. The filled circles mark the primary cycles and the empty the secondary ones. We also over-plotted  the three sequences   identified by \cite{2007ApJ...657..486B} and a fourth line corresponding to the $P_{\rm cyc}/P_{\rm rot}\simeq 5$ ratio of the solar Rieger cycle.  The figure shows that the primary cycles do not follow any particular trend and that their lengths seem to be  uncorrelated with the rotation periods.
The secondary cycles are also uniformly distributed between the Ab and the I sequences and do not follow any particular pattern. Some of the secondary cycles lie beneath the I sequence and their $P_{\rm cyc}/P_{\rm rot}$ ratio is very close to that of the solar Rieger-cycle.

Other authors  \citep[see e.g.][]{1996ApJ...460..848B,2014MNRAS.441.2744V,2016A&A...590A.133O} searched for a correlation between $\frac{P_{\rm cyc}}{P_{\rm rot}}$ and the parameter $\frac{1}{P_{\rm rot}}$ that, according to  \cite{1996ApJ...460..848B}, should be proportional to the dynamo number D.
In these works a linear fit is performed of investigating the relationship between the logarithms of the two quantities.
The slope $m$  extracted from  the linear fit is  the exponent of the power law $P_{\rm cyc}/P_{\rm rot} = (1/P_{\rm rot})^{m}$. If $m$ is close to 1 then no correlation exists between $P_{\rm cyc}$ and $P_{\rm rot}$.    
\cite{1996ApJ...460..848B} found m=0.74 that is in agreement with the value m=0.76 recently found by \cite{2016A&A...590A.133O}.

\begin{figure}
\begin{center}
\includegraphics[width=80mm]{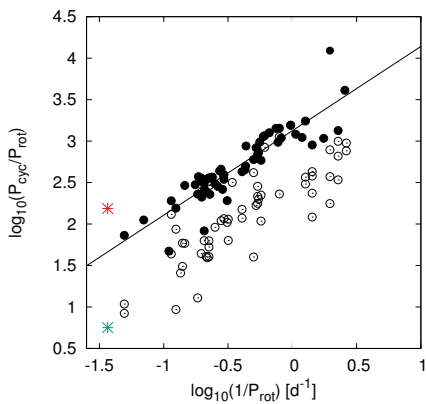}
\caption{The $\frac{P_{\rm cyc}}{P_{\rm rot}}$ ratio vs. the inverse of the stellar rotation period. The filled circles mark the primary cycles and the empty mark the secondary ones. The slope $m=1.02\pm0.06$ of the best-fitting straight line indicates a lack of correlation between $P_{\rm cyc}$ and $P_{\rm rot}$.  Black asterisk: solar Schwabe cycle. Green asterisk: solar Rieger cycle}
\label{pcycprotvida}\end{center}
\end{figure}

In Fig. \ref{pcycprotvida} we plotted $\rm log \frac{P_{\rm cyc}}{P_{\rm rot}}$ vs. $\rm log \frac{1}{P_{\rm rot}}$ for our targets.
We performed a linear fit between the two quantities and we found $m=1.02\pm0.06$ that indicates no correlation between $P_{\rm cyc}$ and $P_{\rm rot}$.
Our results are sligthly higher than those found by  \cite{1996ApJ...460..848B} and \cite{2016A&A...590A.133O}  but are in  close agreement with those found by \cite{2012ARep...56..716S}  and \cite{2016A&A...588A..38L}.

The lack of correlation between $P_{\rm cyc}$ and $P_{\rm rot}$ seen in Fig. \ref{pcycprot} and in Fig. \ref{pcycprotvida} suggests that the dynamo mechanism occurring in our young targets stars is   different from that acting in the older  Mt. Wilson stars selected by \cite{2007ApJ...657..486B}  .

\begin{figure*}
\begin{center}
\includegraphics[width=135mm]{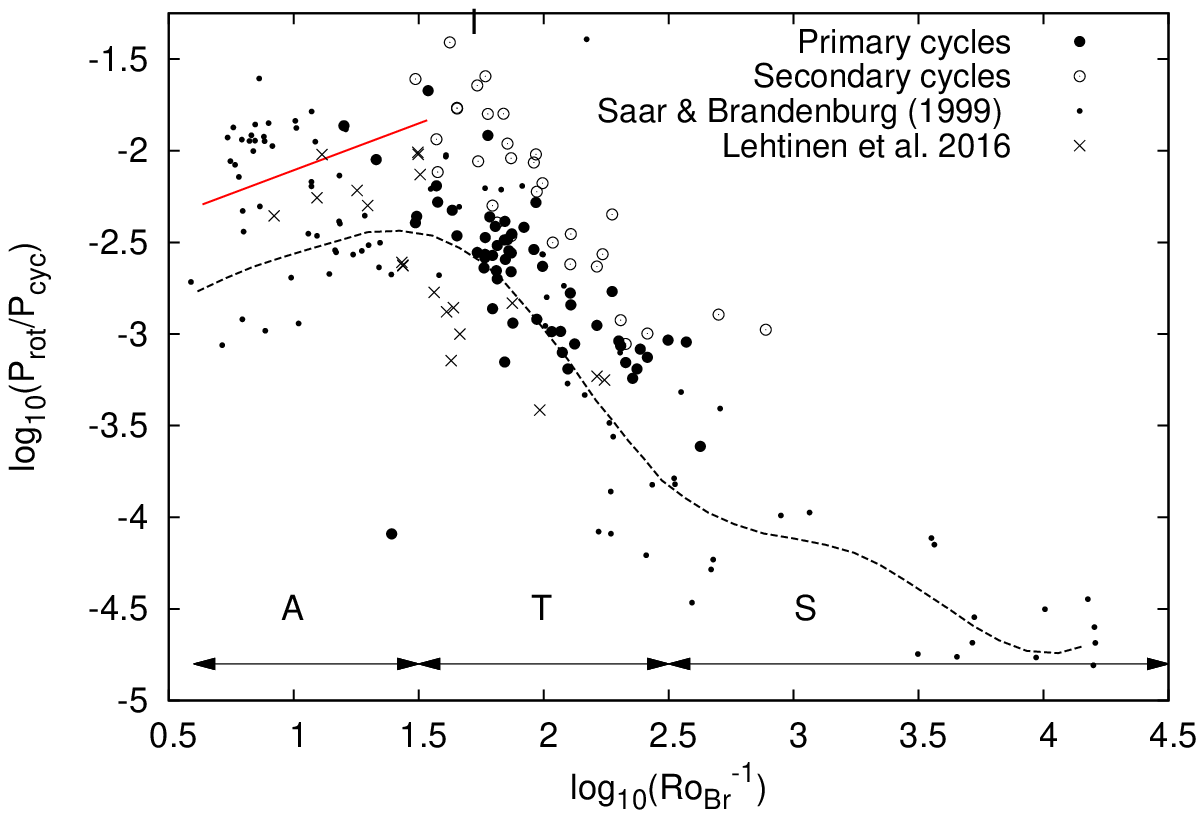}
\caption{ The $\frac{P_{\rm cyc}}{P_{\rm rot}}$ ratio vs. the $Ro_{\rm Br}$ number. The filled bullets mark the primary cycles whereas the empty bullets the secondary ones. The black points and the black crosses mark the results  by \protect\cite{1999ApJ...524..295S} and \protect\cite{2016A&A...588A..38L}, respectively. The red continuous line depicts the Inactive Sequence.  The  black dotted line depicts the curve inferred by \protect\cite{2016A&A...588A..38L} and connects the Active, the Transitional and the Super-Active sequence. The $Ro_{\rm Br}$ ranges corresponding to the three sequences are marked with the black arrows and labelled with A, T, and S, respectively.   }
\label{coriolis}
\end{center}
\end{figure*}

The lack of correlation between $P_{\rm cyc}$ and $P_{\rm rot}$ seen in our targets could be related to their low Rossby number values. 
In fact,  most  of our targets are fast rotating stars ($P_{\rm rot}$ < 10 d) with Ro values falling in the range (0.004-0.17) and $Ro_{\rm Br}$ values  in the range (0.002-0.06).
These ranges are very different from those covered by the Mt. Wilson stars.
\cite{1999ApJ...524..295S} extended the sample of the Mt. Wilson stars with an ensemble of young and fast rotating stars and showed that these stars formed a third branch in the ($\rm log \frac{P_{\rm rot}}{P_{\rm cyc}}$,$\rm log Ro_{\rm Br}^{-1}$) plane. This branch was called the Super Active branch and was clearly  distinct from the Active and Inactive ones and was attributed to a different dynamo mechanism.
Recently,  \cite{2016A&A...588A..38L} exploited long-term photometry to measure the cycle length
$P_{\rm cyc}$ of 21 young active stars and found that the Active and the Super-Active branch are connected by a Transitional branch.
In Fig. \ref{coriolis} we plotted the $\rm log \frac{P_{\rm rot}}{P_{\rm cyc}}$ values vs. the $\rm log Ro_{Br}^{-1}$ values inferred for our stars. In the plot we reported also the data coming from  
\cite{1999ApJ...524..295S} and from \cite{2016A&A...588A..38L}. The  primary cycles of our stars are in good agreement with the Transitional Sequence as defined by \cite{2016A&A...588A..38L}.
Note that also  \cite{2016A&A...588A..38L} did not find any correlation between $P_{\rm cyc}$ and $P_{\rm rot}$ in their targets.
Hence, we can conclude that in stars belonging to the Transitional Branch, the cycle length $P_{\rm cyc}$ and the rotation period $P_{\rm rot}$ are uncorrelated.

\subsection{Relationship between  ${P_{\rm cyc}}$  and global stellar parameters}
We investigated how the cycle lengths $P_{\rm cyc}$ are related to the global stellar parameters of our targets. 
We evaluated the degree of the correlation between $P_{\rm cyc}$ and a given parameter by computing the Spearman rank-order correlation coefficient ($r_S$) \citep{1992nrca.book.....P}.
This coefficient is a non-parametric measure of the monotonicity of the relationship between two datasets.   


A value  of $r_S$ close to 0 implies that the two datasets are poorly correlated, whereas the closer $r_S$ is to $\pm1$ the stronger the monotonic relationship between the two variables.

The statistical significance of a given value  $r_S=r_0$  can be evaluated by computing the two sided p-value  i.e. the probability $(P(|r_S| \ge |r_0|)$  under the null hypothesys that two invesigated datasets are non-monotonically correlated \citep{1992nrca.book.....P}


In Table \ref{spearpcyc} we reported the Spearman correlation coefficients between the length  $P_{\rm cyc}$ of the primary cycles and the different stellar parameters. The corresponding p-values are alo reported.
The values of $|r_S|$ are very close to 0 in all the cases indicating a poor correlation between $P_{\rm cyc}$ and the different variables.

\subsection{Relationship between <IQR> and global stellar parameters}

As remarked in Sec. 2, the IQR index is proportional to the  amplitude of the rotational modulation signature and is related to the non-axisymmetric part of the spot distribution. 
The amplitude of rotational modulation   is a widely used  activity index and can be regarded as a robust proxy of the surface magnetic activity.
\cite{2013JPhCS.440a2020G} demonstrated that the variance of the Total Solar Irradiance, computed on a 60-d sliding window, closely mimics the variations of the 10.7 cm radio flux that in turn is a good indicator of the solar magnetic activity \citep[see][for details]{2014JApA...35....1B}.
We computed the Spearman coefficient between <IQR> and different stellar parameters in order to investigate how the mean level of the stellar surface magnetic activity is linked to the stellar properties.
In Table \ref{speariqr} we reported the values of $r_{\rm S}$ for the different parameters. In this case, $r_{\rm S}$ values are significantly higher then those computed for $P_{\rm cyc}$. A good correlation ($|r_{\rm S}|\simeq 0.5$ is seen between <IQR> and $T_{\rm eff}$, $\tau_{\rm C}$, $\tau_{\rm diff}$, $R_{\rm C}$ and $D_{\rm {\rm N}}$. The sign of correlation is positive for $\tau_{\rm C}$, $\tau_{\rm diff}$ and $D_{\rm N}$ and negative for $T_{\rm eff}$ and $R_{\rm C}$. 

\begin{figure*}
\begin{center}
\includegraphics[width=160mm]{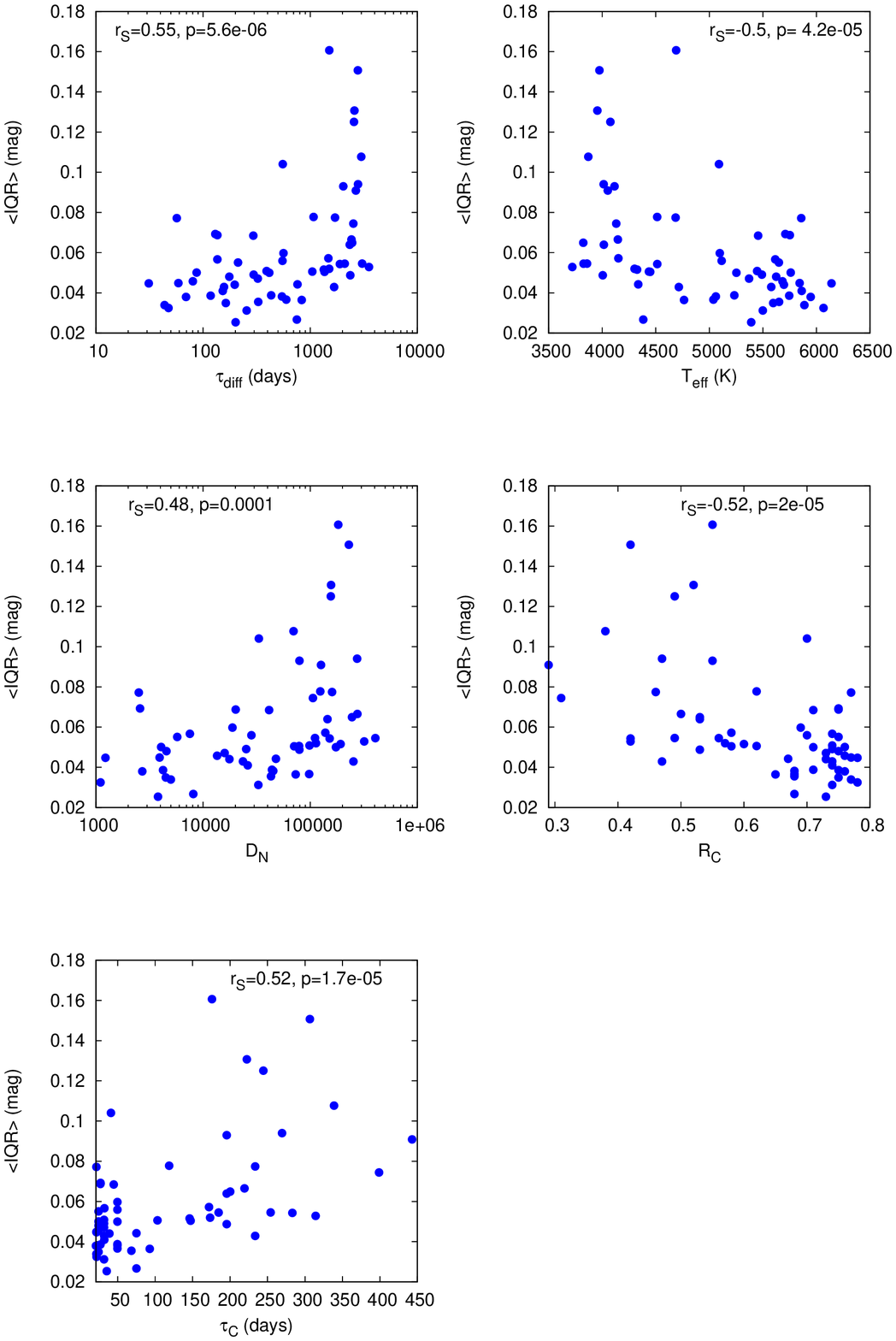}
\caption
{In the different panels we reported <IQR> vs. the parameters $\tau_{\rm diff}$, $T_{\rm eff}$, $D_{\rm N}$, $R_{\rm C}$ and $\tau_{\rm C}$, respectively. In each panel we also reported the Spearman coefficient $r_{\rm S}$ and the corresponding two-sided p-value.  }
\label{iqrpars}
\end{center}
\end{figure*}

 In Fig. \ref{iqrpars} we reported <IQR> vs. the different stellar parameters for which a significant correlation is seen ($r_{\rm S} \simeq 0.5$ ).
Despite the scatter in the data, the different plots  clearly show that <IQR> tends to increase with $\tau_{\rm diff}$, $\tau_{\rm C}$ and $T_{\rm eff}$ and it has a negative correlation with $T_{\rm eff}$  and $R_{\rm C}$ .

\begin{figure}
\begin{center}
\includegraphics[width=80mm]{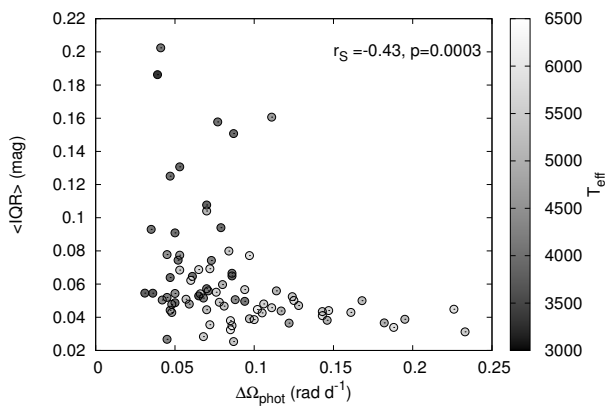}
\caption{<IQR> vs. $\Delta\Omega_{\rm phot}$. The points are color coded according the stellar effective temperature $T_{\rm eff}$.}
\label{iqrdo}
\end{center}
\end{figure}

A weaker but still significant correlation is also seen between <IQR> and the photometric shear $\Delta\Omega_{\rm phot}$.  In Fig. \ref{iqrdo} we plotted <IQR> vs. $\Delta\Omega_{\rm phot}$.  The different stars are color-coded according to their effective temperature $T_{\rm eff}$. The plot shows  that:
\begin{itemize} 
\item{$\Delta\Omega_{\rm phot}$ tends to increase with temperature (as demonstrated in Paper I)}
\item{ the sign of the correlation between <IQR> and $\Delta\Omega_{\rm phot}$ and <IQR>  is negative ($r_S=-0.43$). This means that the lower $\Delta\Omega_{\rm phot}$ the higher <IQR>.}
\end{itemize}
These results are in very good agreement with  the theoretical models developed by \cite{2011MNRAS.411.1059K}. In fact, these models predict that even a small surface differential rotation is very efficient for dynamos in M-type stars, it is less efficient in K- and G- type and even a strong differential rotation is inefficient in F type stars.

\begin{table}
\caption{Spearman correlation coefficient between $P_{\rm cyc}$ and different stellar parameters. The parameters are ordered by decreasing $|r_S|$.}
\label{spearpcyc}
\centering
\begin{tabular}{lll}
\hline
Parameter & $r_S$    & p     \\
\hline
$P_{\rm rot}$ & 0.16 & 0.25 \\
$\Delta\Omega_{\rm phot}$ &-0.10 & 0.48\\
$D_{\rm N}$&-0.10& 0.47\\
$\tau_{\rm C}$& 0.04& 0.79\\
$R_{\rm C}$& -0.04& 0.76\\
Ro& 0.03 &0.84\\
$T_{\rm eff}$& -0.03& 0.86\\
age &0.02 & 0.86 \\
$\tau_{\rm diff}$ &0.01& 0.95\\
\hline
\end{tabular}
\end{table}

\begin{table}
\caption{Spearman correlation coefficient $r_S$ between <IQR> and different stellar parameters. The parameters are ordered by decreasing $|r_S|$.} 
\label{speariqr}
\centering
\begin{tabular}{lll}
\hline
Parameter & $r_S'$    & p     \\
\hline
$\tau_{\rm diff}$ &0.55& 5.6e-06\\
$\tau_{\rm C}$& 0.52& 1.7e-05\\
$R_{\rm C}$& -0.52& 2e-05\\
$T_{\rm eff}$ & -0.50& 4.2e-05\\
$D_{\rm N}$&0.48& 0.0001\\
$\Delta\Omega_{\rm phot}$ &-0.43 & 0.0003\\
age &-0.39 & 0.001 \\
Ro& -0.35 &0.0061\\
$P_{\rm rot}$ & 0.29 & 0.018 \\
$P_{\rm cyc}$ &0.08 & 0.56\\
\hline
\end{tabular}
\end{table}

\subsection{Evolution of  $P_{\rm cyc}$  and <IQR> with the stellar age}
\begin{figure}
\begin{center}
\includegraphics[width=80mm]{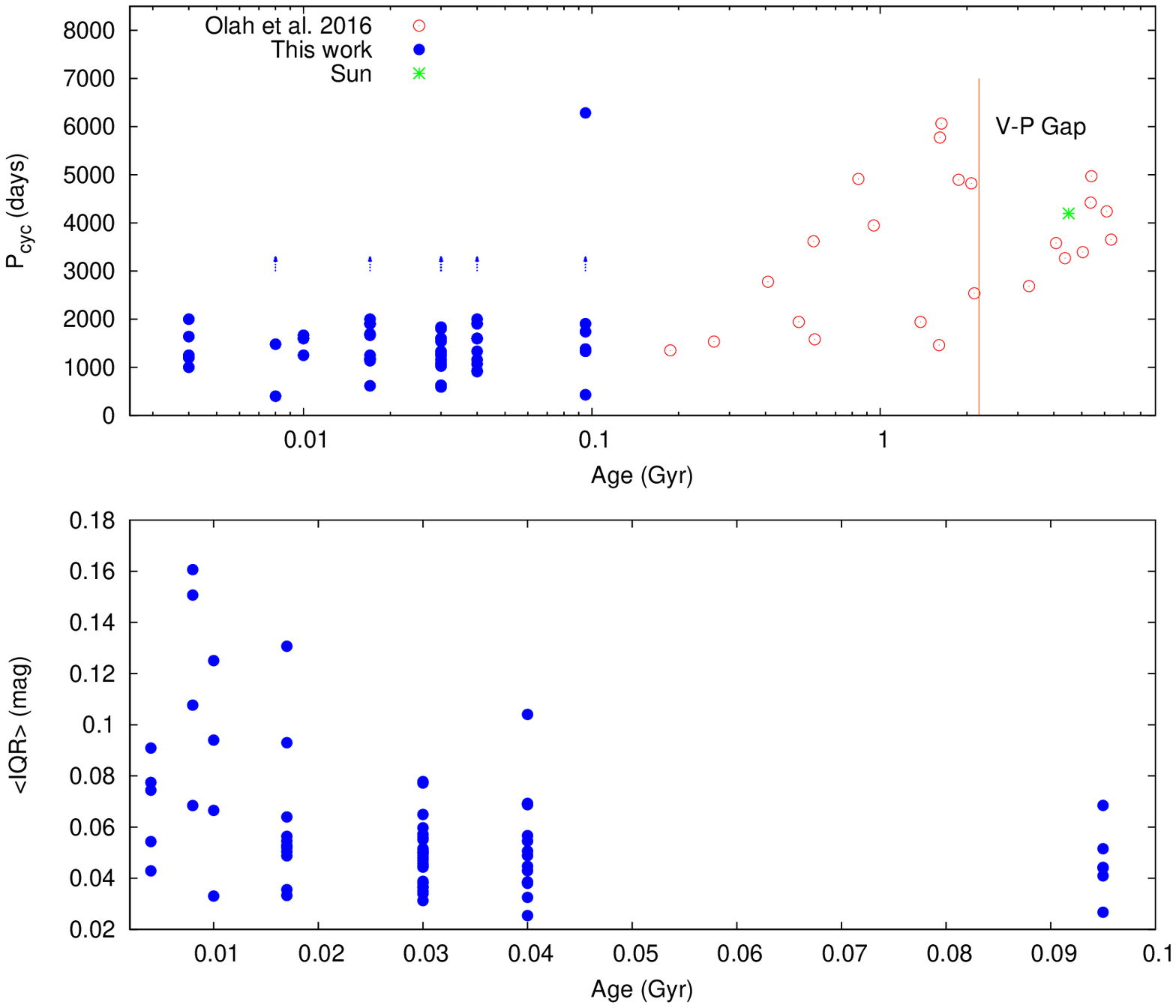}
\caption{  Top panel. $P_{\rm cyc}$ vs. the stellar age. The blue filled circles refer to the primary cycles detected in the present work and the red empty circles to the data reported by \protect\cite{2016A&A...590A.133O}. The blue arrows mark the stars for which the visual inspection of time-series revealed  $P_{\rm cyc} \ge 3000~d$. Bottom panel. The activity index <IQR> vs. the stellar age.}
\label{pcycage}
\end{center}
\end{figure}

\cite{2016A&A...590A.133O} investigated how the  $\frac{P{\rm cyc}}{P_{\rm rot}}$ ratio depends on the stellar age. They found that   $\frac{P_{\rm cyc}}{P_{\rm rot}}$ exhibits a large scatter in the younger and more active stars of their sample whereas  the older stars have about the same value. The age at which $\frac{P_{\rm cyc}}{P_{\rm rot}}$ becomes  constant  is, according to their work,  at about  2.2 Gyr, i.e. the age of the Vaughan-Preston gap \citep{1980PASP...92..385V}.
The sample of stars analyzed by \cite{2016A&A...590A.133O} covers the age range 200-6200 Myr. Our work permits us to extend the age range studied by \cite{2016A&A...590A.133O} and to investigate  the transition age between the PMS and the MS phase.
Our analysis is slightly different from that performed by \cite{2016A&A...590A.133O}. We decided to investigate the relationship between $P_{\rm cyc}$ and the stellar age  instead of the relationship between $\frac{P_{\rm cyc}}{P_{\rm rot}}$ and the stellar age. In fact, the trend of the between $\frac{P_{\rm cyc}}{P_{\rm rot}}$ ratio, in the age range covered by our targets, could be  dominated by the complex $P_{\rm rot}$ evolution seen in the PMS stars \citep[see e.g.][and references  therein]{2015A&A...584A..30L}. 
In the top panel of Fig. \ref{pcycage}, we reported the length of the primary cycles $P_{\rm cyc}$  vs. the stellar age.
The blue filled symbols mark the cycles identified in the present work and the red empty symbols mark those found by \cite{2016A&A...590A.133O}. 
The picture shows that  $P_{\rm cyc}$ is about constant in the age range $4-300~\rm Myr$. After 300 Myr, the cycle lengths are more scattered and seem to increase with the stellar age. After 2.2 Gyr, that is the age corresponding to the Vaughan-Preston gap, the scatter in $P_{\rm cyc}$  decreases as described by \cite{2016A&A...590A.133O}.
Note that, although our data cover an interval of about 3000 d which could prevent the detection of longer cycles, we observed long-term trends suggesting cycles longer than 3000 d just in 11 of the 90 investigated targets. For this reason, we are confident that our data are not seriously affected by a selection bias and that the age range investigated here is really, on average, characterized by cycles shorter than those observed in the older Mt. Wilson stars.
In the bottom panel of Fig. \ref{pcycage} we plotted <IQR> vs. the stellar age. The picture shows that <IQR> tends to decrease with the stellar age as also indicated by the Spearman coefficient $r_{\rm S}=0.39$. This result is in agreement with all the works that studied the evolution of the level of magnetic activity vs. age \citep[see e.g.][and references therein]{2017ApJ...835...61Z,2016A&A...590A.133O}

\subsection{Butterfly diagrams}
\begin{figure*}
\begin{center}
\includegraphics[width=160mm]{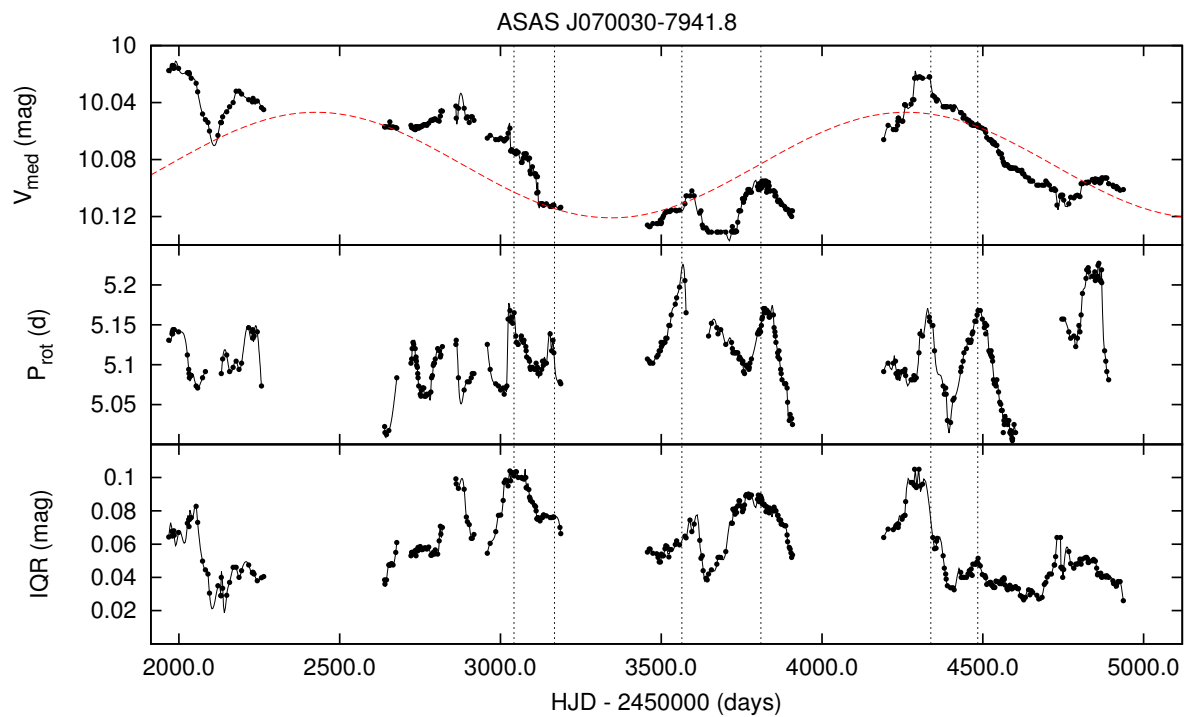}
\caption{Top panel: the $V_{\rm med}$ time-series for the star ASAS J070030-7941.8.  Mid panel; the $P_{\rm rot}$ time-series. Bottom panel: the IQR time-series. A primary cycle with length $P_{\rm cyc}=1834~d$ was detected in the $V_{\rm med}$ time-series and the sinusoid best-fitting the data was over-plotted on it (red dashed line). The $P_{\rm rot}$ and the IQR time-series exhibit cycles (dark continuous lines) shorter than the primary one. The black dotted lines mark the times at which maxima occur in the $P_{\rm rot}$ time-series. }
\label{butterfly1}
\end{center}
\end{figure*}

\begin{figure*}
\begin{center}
\includegraphics[width=160mm]{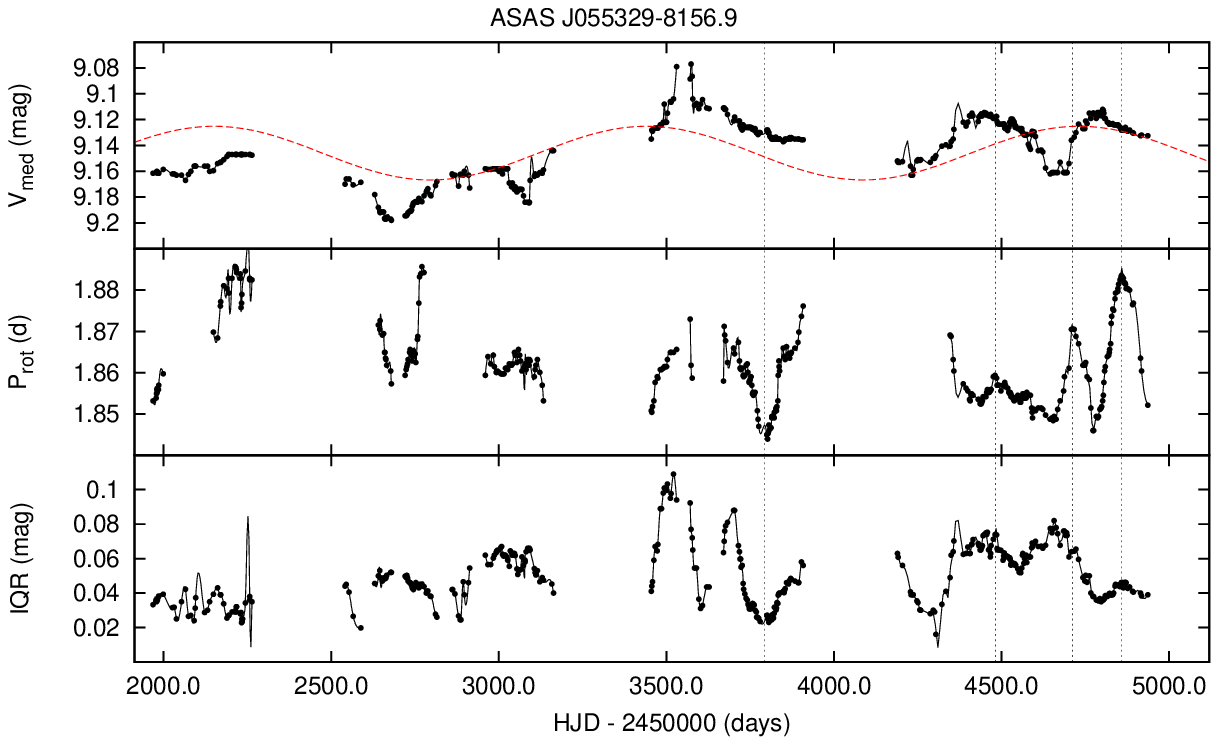}
\caption{ Top panel: the $V_{\rm med}$ time-series for the star ASAS J055329-8156.9.  Mid panel; the $P_{\rm rot}$ time-series. Bottom panel: the IQR time-series. A primary cycle with length $P_{\rm cyc}=1290~d$ was detected in the $V_{\rm med}$ time-series and the sinusoid best-fitting the data was over-plotted on it (red dashed line). The $P_{\rm rot}$ and the IQR time-series exhibit cycles (dark continuous lines) shorter than the primary one. The black dotted lines mark the times at which maxima and minima occur in the $P_{\rm rot}$ time-series. }
\label{butterfly2}
\end{center}
\end{figure*}

In the Sun, the latitudes at which ARs occur change vs. time according to the so-called ``butterfly-diagram''. At the beginning of a solar cycle, ARs emerge  in belts located at intermediate latitudes ($\pm 35 \deg$).  As the cycle progresses, the ARs formation regions migrate toward the equator. At the beginning of the next cycle, the ARs form again at intermediate latitudes. 
The solar SDR combined with the ARs migration determines a decrease in the rotation period detectable along the 11-yr cycle because the ARs located at intermediate latitudes rotate slower than those located at the equator.
As remarked in Paper I and in Sec. 2.2,  the Activity Index $P_{\rm rot}$  of  a given star can therefore be regarded as a tracer of the ARs migration in latitude.
This Activity Index cannot give information about the exact latitudes  at which ARs occur but can give insight about the nature of stellar magnetic field. In fact, in a star with a solar-like dynamo, we expect that the trend of $P_{\rm rot}$ with time mimics that of the solar butterfly diagram i.e. that, during a given cycle, $P_{\rm rot}$ decreases  in time and that it rises rapidly to higher values at the beginning of the next cycle.
\cite{2003A&A...409.1017M}, for instance, analyzed the trend of $P_{\rm rot}$ in six young solar analogues. They noticed that three of these stars follow a solar behaviour whereas the other three show an anti-solar behaviour with $P_{\rm rot}$ increasing during the cycle. 
The $P_{\rm rot}$ time-series of our target stars have in general  less points than the $V_{\rm med}$ and IQR time-series. Indeed, as remarked in Paper I, $P_{\rm rot}$ is detectable in a given time interval only if ARs maintain  a stable configuration. 
In Fig. \ref{butterfly1} and \ref{butterfly2} we reported the $P_{\rm rot}$ time-series for the stars ASAS J070030-7941.8 and ASAS J055329-8156.9 that are the targets with the highest number of $P_{\rm rot}$ determinations. The $V_{\rm med }$ time-series and the IQR time-series of the two stars are also reported for comparison. The stars have two primary cycles with length $P_{\rm cyc}~= 1834~d$ and $P_{\rm cyc}~= 1290~d$, respectively. The two sinusoids best-fitting the data and corresponding to the primary cycles are over-plotted on the $V_{\rm med}$ time-series.
The visual inspection of the pictures shows that in both stars:
\begin{itemize}
\item{The $P_{\rm rot}$ index oscillates with a  cycle shorter than the primary cycle; hence the ARs latitude migration occurs in a time-scale shorter than the length of the primary cycle; }
\item{in some intervals, the trend of $P_{\rm rot}$ vs time mimics the trends of $V_{\rm med}$ and IQR and the maxima and minima  of $P_{\rm rot}$ correspond to local maxima and minima  of the $V_{\rm med}$ and $IQR$ time-series; in other intervals   the maxima and minima of $P_{\rm rot}$ are uncorrelated with those of the $V_{\rm med}$ and IQR time-series. This implies that in some time intervals the changes in AR latitudes are  correlated with the variations in the ARs areas  and in the mean level of magnetic activity whereas in some other intervals there is no correlation;   }
\item{the $P_{\rm rot}$ variations with time are very different from those occurring in the Sun. Rise patterns are followed by decreasing patterns. This implies that ARs first migrate from regions with shorter rotation periods to regions with higher rotation periods and then migrate in the opposite sense. In the Sun the migration occurs in only one sense i.e. from the intermediate latitudes to the equator. }
\end{itemize}
Similar trends are observed in all the targets analyzed here.

The patterns seen in the $P_{\rm rot}$ time-series of our targets can be regarded as a further evidence that the dynamo acting in these stars is very different from that acting in the Sun. 

\section{Conclusions}
\label{sec:summary}
In the present work we searched for activity cycles in stars belonging to young loose stellar associations. 
We analyzed the long-term time-series of three different activity indexes and we were able to detect activity cycles in 67 stars and to measure their length $P_{\rm cyc}$.
We investigated how $P_{\rm cyc}$ is correlated with global stellar parameters by computing the Spearman coefficient $r_{\rm S
}$ between $P_{\rm cyc}$ and the different parameters. 
We investigated how activity cycles evolve with the stellar age.
In particular, our work  extends the age range covered by the Mt Wilson stars whose properties have been recently  reanalyzed by \cite{2016A&A...590A.133O}.
Our analysis led to the following results:
\begin{enumerate}
\item{most of the analyzed stars  show multiple and complex cycles according to the results found by \cite{2016A&A...590A.133O} for young and active stars; }
\item{some of the detected secondary cycles have a $\frac{P_{\rm cyc}}{P_{\rm rot}}$ ratio similar to that of  the solar Rieger-cycles ($\frac{P_{\rm cyc}}{P_{\rm rot}}\simeq~5$); }
\item{the location of our targets in the $\rm log \frac{P_{\rm cyc}}{P_{\rm rot}}-\rm log Ro_{\rm Br}^{-1}$ plane is in good agreement with the Transitional branch as defined by \cite{2016A&A...588A..38L};}
\item{the cycle length $P_{\rm cyc}$ is uncorrelated with the stellar rotation period $P_{\rm rot}$ in our targets; this result confirms that found by \cite{2016A&A...588A..38L} for stars belonging to the Transitional Branch;}
\item{$P_{\rm cyc}$ is essentially uncorrelated with global stellar parameters in the age range we investigated;} 
\item{the activity index <IQR>, that can be regarded as a proxy of the magnetic surface activity level, is positively correlated with $\tau_{\rm C}$, $\tau_{\rm diff}$, $D_{\rm N}$ and negatively correlated with $T_{\rm eff}$  and$R_{\rm C}$;  }
\item{   in agreement with the \cite{2011MNRAS.411.1059K} model, even a small differential rotation is efficient for dynamos in M-type stars, but it becomes less efficient or completely inefficient at increasing $T_{\rm eff}$;}
\item{the analysis of the butterfly diagrams of our target stars shows that:}
\begin{itemize}
\item{the ARs migration in latitude occurs over a time-scale shorter than the primary cycle;}
\item{the latitudes at which ARs emerge seem to oscillate from a maximum to a minimum latitude and vice-versa. This is very different from the solar behaviour where the spots migrate only  from intermediate latitudes to the equator;}
\end{itemize}
\item{ we merged our data with those analyzed by \cite{2016A&A...590A.133O} and we studied the trend of  $P_{\rm cyc}$ vs. the stellar age in the range  (0.004-9 Gyr); we found that $P_{\rm cyc}$ is about constant and does not show significant correlation with the stellar age in the range ($4-300\rm Myr$); after 300 Myr $P_{\rm cyc}$ values are quite scattered and  tend to increase with the stellar age; after 2.2 Gyr, $P_{\rm cyc}$ values tend to be less scattered and seem to converge to the solar value as described by \cite{2016A&A...590A.133O}; }
\item{the activity index <IQR> decreases with the stellar age in the age range 4-95 Myr;}
\item{merging the ASAS time-series of the star AB Dor A with the photometric data  from previous works and we detected a cycle with length $P_{\rm cyc} = 16.78 \pm 2 \rm yr$ and shorter secondary cycles with lengths of 400 d, 190 d, and 90 d. }
\end{enumerate}

\begin{acknowledgements} 
The authors are grateful to the referee Lauri Jetsu for helpful comments and suggestions. 
 \end{acknowledgements}
\bibliographystyle{aa}
\bibliography{SDRref}

\begin{appendix}
\section{Effect of the sliding-window algorithm on a periodic signal.}
\label{signals}
\begin{figure}
\begin{center}
\includegraphics[width=80mm]{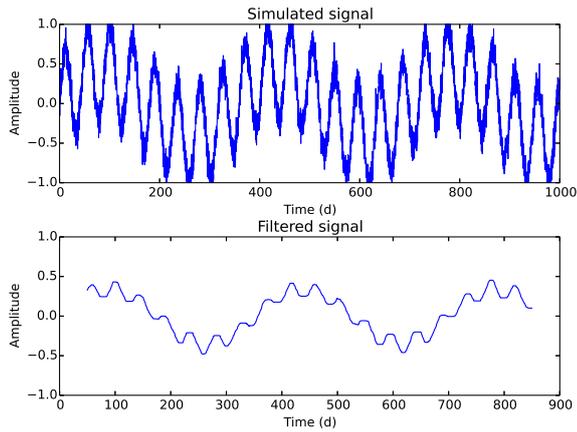}
\caption{ \textbf{Top panel}. A simulated time-series obtained by combining two sinusoidal signals with periods $\rm P_1 = 300 \rm d$, $\rm P_2= 45 \rm d$ and amplitudes $\rm A_1=0.4$,  $\rm A_2=0.6$, respectively.
\textbf{Bottom panel}. Filtered time-series. The segmentation algorithm attenuates but does not suppress the signal with P = 45 d.  }
\label{signal1}
\end{center}
\end{figure}

\begin{figure}
\begin{center}
\includegraphics[width=80mm]{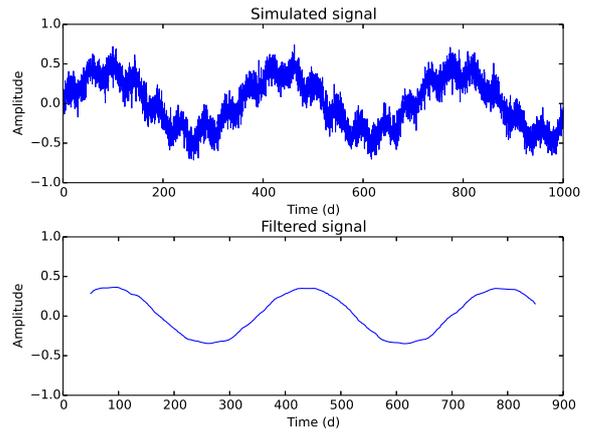}
\caption{ \textbf{Top panel}. A simulated time-series obtained by combining two period signals with periods $\rm P_1 = 300 \rm d$, $\rm P_2= 45 \rm d$ and amplitudes $\rm A_1=0.4$,  $\rm A_2=0.1$, respectively.
\textbf{Bottom panel.} Filtered time-series. The 45-d is completely suppressed in this case.  }
\label{signal2}
\end{center}
\end{figure}

The segmentation procedure used to derive activity indexes is equivalent to a low-pass filter. This preserves the signals with a frequency lower than a certain cutoff frequency and attenuates signals with frequencies higher than the cutoff frequency.
In our case, the use of a sliding-window with length T=100 d attenuates  signals with periods shorter than 100-d but, in some cases,  does not completely suppress them.
We performed different tests by simulating sinusoidal signals with different amplitudes and periods and by processing them with our segmentation algorithm. In the top panel of Fig. \ref{signal1} we plotted a simulated time-series obtained by combining two sinusoidal signals with periods $\rm P_1 = 300 \rm d$, $\rm P_2 = 45 \rm d$ and amplitudes $\rm A_1 = 0.4$,  $\rm A_2 = 0.6$, respectively. A white gaussian noise with variance $\sigma=0.1$ was added to the simulated data.  In the bottom panel,    we plotted the filtered signal obtained after processing the simulated time-series with our segmentation algorithm. The filter has attenuated but not suppressed the 45-d signal. 
In Fig. \ref{signal2} we reported a similar test obtained by combining two period signals $\rm P_1 = 300 \rm d$, $\rm P_2= 45 \rm d$ and amplitudes $\rm A_1=0.4$,  $\rm A_2=0.1$. In this second case the signal with $\rm P=45 \rm d$ is completely suppressed.
\end{appendix}

\label{lastpage}

\end{document}